\documentclass[11pt, a4paper]{article}

\usepackage[utf8]{inputenc}
\usepackage[T1]{fontenc}
\usepackage[english]{babel}
\usepackage{amsmath, amssymb}
\usepackage{booktabs}
\usepackage{geometry}
\usepackage{hyperref}
\usepackage{parskip}
\usepackage{array}
\usepackage{graphicx}
\usepackage{caption}
\usepackage[numbers]{natbib}
\usepackage{subcaption}
\usepackage{xcolor}
\usepackage{siunitx}

\hypersetup{colorlinks=true, linkcolor=black, citecolor=black, urlcolor=blue}

\title{\textbf{Network topology shapes consensus and vulnerability in LLM populations}}
\author{
  Roberto Passaro\textsuperscript{a,b},
  Ariel Flint\textsuperscript{c},
  Romualdo Pastor-Satorras\textsuperscript{d},\\
  Andrea Baronchelli\textsuperscript{c,*}, and Luca Maria Aiello\textsuperscript{a,b},\\[1.5ex]
  {\small \textsuperscript{a}IT University of Copenhagen, Denmark.}\\
  {\small \textsuperscript{b}Pioneer Centre for AI, Copenhagen, Denmark.}\\
  {\small \textsuperscript{c}Department of Mathematics, City St George's, University of London, UK.}\\
  {\small \textsuperscript{d}Departament de F{\'\i}sica, Universitat Polit{\`e}cnica de Catalunya, Barcelona, Spain.}\\[1ex]
  {\small \textsuperscript{*}Corresponding author: \url{a.baronchelli.work@gmail.com}}
}

\date{October 2026}

\begin{document}
\maketitle

\begin{abstract}
As large language models (LLMs) are deployed in interacting populations, their collective behaviour depends not only on the models but also on who interacts with whom. Studies of convention formation in LLM populations have mainly tested well-mixed interactions. We place policies extracted from four LLMs on fully connected, random regular, and heterogeneous networks, with $10^3$ to $10^5$ agents playing the naming game. Topology affects how quickly consensus emerges, sometimes whether it emerges at all, and, in rare cases, which convention wins. Sparsest networks slow consensus up to twelvefold. Yet, by preserving correlations between persistent neighbours, they restore consensus in some cases that fail to converge under well-mixed interactions. Networks also govern vulnerability to adversarial attack. We measure the critical mass of committed agents needed to overturn an established convention. Tipping thresholds follow the same ordering across models and conditions: hubs $\leq$ random $\leq$ low-degree nodes. On the most heterogeneous topology, median critical masses are 3\%, 16\%, and 35\%, respectively; targeting hubs requires one-sixth as many committed agents as random placement. For most models, networked populations are easier to tip than well-mixed ones, so well-mixed estimates can overstate robustness. Assessing the safety of multi-agent AI requires mapping who interacts with whom.
\end{abstract}



\section{Introduction}

AI agents are moving from isolated tools to interacting populations~\cite{bengio2026international, hammond2025multi}. Multi-agent systems based on large language models (LLMs) already operate in finance~\cite{ferreira2021artificial,sun2023reinforcement}, defence~\cite{black2024strategic}, social media~\cite{schroeder2026malicious}, and personal assistance~\cite{li2024personal}. The collective behaviour of these populations is becoming directly observable at multiple scales. In Emergence World, a continuously running multi-agent environment, groups of dozens of agents have been observed cooperating to create and maintain shared institutions~\cite{akkil2026emergence}. In July 2026, about $1200$ agents at OpenAI exchanged more than $\num{70000}$ messages and files and coordinated collective projects to tamper with the scorer of the benchmark they were working on, resulting in a successful cyberattack on Hugging Face's digital infrastructure~\cite{greenblatt2026brief}. On Moltbook, an agent-only social platform, hundreds of thousands of agents organise into communities with a highly hierarchical interaction structure~\cite{price2026let, li2026rise}.

Many interactions in these systems aim to reach agreement on \emph{conventions}: shared behaviours that help achieve common goals without a central coordinating authority~\cite{wittgenstein1953philosophical,lewis2008convention, young1993evolution, bicchieri2005grammar}. The centrality of conventions has motivated experimental work on their emergence in AI multi-agent systems, often based on minimal models that capture the core coordination mechanisms. One such model is the \emph{naming game}, in which pairs of agents try to coordinate on a name~\cite{steels1995self}. Theoretical studies~\cite{baronchelli2006sharp, castellano2009statistical, baronchelli2018emergence} and laboratory experiments with humans~\cite{centola2015spontaneous, centola2018experimental, ehret2022group, andreoni2021predicting} have shown that, in the naming game, conventions can emerge spontaneously from repeated local interactions.

Recent simulations with populations of LLM-based agents have revealed similar patterns. AI agents playing the naming game can reach global conventions through local interactions, although these interactions can produce collective biases absent in individual agents~\cite{ashery2025emergent}. The strength and form of these biases vary with population size in nonlinear, model-dependent ways~\cite{flint2026group}. Established conventions can also be overturned. A committed minority, a group of agents that always promote an alternative, can tip the whole population once it exceeds a critical fraction~\cite{xie2011social, centola2018experimental}. This critical mass is key to the stability of these systems because it measures a social attack surface: the number of adversarial agents needed to overturn a convention a population has already settled on~\cite{flint2026indirect}. Critical mass has been measured at $25\%$ in experiments with human groups~\cite{centola2018experimental}, ranges from $10\%$ to $40\%$ in theoretical models depending on their assumptions~\cite{xie2011social, niu2017impact}, and spans $2\%$ to $67\%$ in AI populations depending on the underlying LLM~\cite{ashery2025emergent}.

All these studies of coordination in AI multi-agent systems share one simplification: the population is \emph{well mixed}, so every agent can interact with every other~\cite{ashery2025emergent, flint2026group, flint2026indirect}. This is rarely true in real-world deployments, where the communication network among agents can be shaped by top-down architectural design~\cite{qian2025scaling}, adaptive partner discovery to optimise task delegation~\cite{ehtesham2025survey}, and uneven activity levels across agents~\cite{price2026let}.

In the classical naming game, interaction structure has well-known effects. On low-dimensional lattices, consensus proceeds through the slow coarsening of local domains~\cite{baronchelli2006topology}. On heterogeneous networks, high-degree nodes act as spreaders of conventions~\cite{dall2006nonequilibrium}, and on sparse random networks the tipping point shifts~\cite{zhang2012analytic}. In human experiments, a global convention emerged in homogeneously mixing populations but not on one-dimensional lattices or random  networks of degree $4$~\cite{centola2015spontaneous}. LLM-derived policies, however, are biased, rely on bounded memory, and exhibit asymmetric critical masses and mixed states absent from the classical model~\cite{ashery2025emergent, flint2026group}. To our knowledge, networked LLM populations playing the naming game have been studied only in groups of $16$ agents~\cite{mehdizadeh2026exploring}, and the critical mass of an LLM population has never been measured on a network.

Here, we study how network topology shapes the formation of conventions in LLM populations and their robustness to committed minorities. We use the policies extracted from four LLMs by~\citet{flint2026group}, which allow simulations of up to $10^5$ agents. We place them on four topologies: a fully connected network, equivalent to the well-mixed setting; two uncorrelated networks with power-law degree distributions~\cite{barabasi1999emergence}, one heterogeneous and one nearly homogeneous; and a random regular network. 
We find that networks change how fast consensus is reached and, for some word pairs, which convention wins and whether it is reached at all.
For three of the four models, the fully connected network is the fastest and the random regular network the slowest. In two cases, the networks with the lowest mean degree also reverse which convention wins. Networks homogeneous and low average degree restore restore consensus for one of the two word pairs that fail to converge when well-mixed, 
requiring the presence of dynamical correlations between neighbouring agents. Hubs reach a given level of coordination earlier than low-degree nodes, spreading the consensus to the rest of the network. Yet they are the cheapest place from which to overturn a convention. In every condition we tested, the critical fraction is lowest when the committed agents occupy the hubs and highest when they occupy the low-degree nodes. Overall, our work extends the study of LLM populations from well-mixed to networked settings, and identifies the structure of interactions as a key factor in how the collective behaviour of multi-agent AI systems emerges and how it can be exposed to adversarial attack.

\section{Methods}

\subsection{Naming game and agent model}

We study the emergence of linguistic conventions using the minimal naming game, in which pairs of agents repeatedly interact and attempt to coordinate on one of $W = 2$ competing conventions~\cite{baronchelli2006sharp, ashery2025emergent}. At each interaction, an agent produces a candidate word informed by its memory of the outcomes of its $H$ most recent interactions. Memories are empty at the start of a simulation. When two interacting agents produce the same word, the interaction is scored as successful and both agents' scores increase; a mismatch is scored as a failure, with the reward for success exceeding the penalty for failure. This incentivises local coordination between interacting pairs, but agents are never given an explicit incentive towards population-wide consensus.

In this study, an agent's behaviour is represented by a probabilistic policy over the $W$ candidate words, conditioned on its current memory state. Rather than querying a live language model at every interaction, which is computationally prohibitive at the population sizes explored here, we reuse policies extracted from four instruction-tuned large language models by \citet{flint2026group}: Meta Llama 3.1 70B Instruct (denoted as Llama), OpenAI GPT-4o (GPT), Qwen QwQ-32B (QwQ), and Microsoft Phi-4 (Phi). In  \citet{flint2026group}, each model was prompted with a textual description of a given memory state, and its output probabilities over the two candidate words were extracted to give a complete stochastic policy for every reachable memory state. This representation was validated against direct LLM-based simulation and reproduces its dynamics closely. We adopt the eleven word pairs used in that study, drawn from the CrowS-Pairs stereotype benchmark \cite{nangia2020crows} and spanning gender, age, sexual orientation, race and socioeconomic stereotypes: \{her, his\}, \{straight, gay\}, \{he, she\}, \{Black, White\}, \{less, more\}, \{old, young\}, \{White, African\}, \{American, Mexican\}, \{short, tall\}, \{man, woman\}, \{husband, wife\}. We adopt the memory window $H = 5$ used there.

The naming game for each populations of $N$ agents lasts for up to $T_{\max} = \num{100000}$ population rounds, where one round comprises $N$ pairwise interactions. A population is judged to have reached consensus once its rolling coordination-success rate exceeds $0.98$; the round at which this occurs is recorded as the convergence time. Populations that do not cross this threshold within $T_{\max}$ are classified as non-converged. We call a pair of words that does not reach convergence a ``mixed-state pair''.

\subsection{Networks and parameters}
\label{sec:networks}

Well-mixed populations lie on a fully connected network (FC): every agent can interact with any other agent in the population. We compare FC with two other topologies: (i) uncorrelated networks with a power-law degree distribution~\cite{barabasi1999emergence}, $P(k) \propto k^{-\gamma}$, generated, using the uncorrelated configuration model (UCM)~\cite{catanzaro2005generation}, at two exponents, $\gamma = 2.5$ (heterogeneous; UCM$_{2.5}$ from now on) and $\gamma = 5.0$ (near-homogeneous; UCM$_{5.0}$), and (ii) a random regular network (RRN), in which every node has the same degree. In UCM, each of the $N$ nodes is assigned a degree drawn from $P(k) \propto k^{-\gamma}$, and restricted to $k \in [k_{\min}, \sqrt{N}]$. The resulting stubs are then joined uniformly at random, rejecting self-loops or repeated edges. The exponent $\gamma$ controls the tail of the degree distribution. At $\gamma = 2.5$ the distribution of the topology is heavy-tailed, and is characterised by the presence of high-degree nodes called \emph{hubs}. Conversely, at $\gamma = 5$, the degree distribution decays rapidly, leading to a more homogeneous topology.
Table~\ref{tab:networks} reports the resulting degree statistics at $N = \num{10000}$. Both UCM configurations share a minimum degree $k_{\min} = 4$, which is also the fixed degree of the RRN.

\begin{table}[t!]
\centering
\caption{Network topologies and their degree statistics at $N = 10000$.}
\label{tab:networks}
\begin{tabular}{llll}
\toprule
Topology & Full name & Degree distribution & $\langle k \rangle$ \\
\midrule
FC               & Fully Connected                        & regular, $k = N - 1$     & $N - 1$ \\
UCM$_{2.5}$ & Uncorrelated Configuration Model       & power-law, 
$\gamma = 2.5$ & 8.64 \\
UCM$_{5.0}$ & Uncorrelated Configuration Model       & power-law, 
$\gamma = 5.0$ & 4.76 \\
RRN              & Random Regular Network                 & regular, $k=4$           & 4.00 \\
\bottomrule
\end{tabular}
\end{table}

We test four population sizes, $N \in \{\num{1000}, \num{5000}, \num{10000}, \num{100000}\}$, keeping all other parameters fixed. At each interaction we consider two rules for selecting which pair of agents interacts: an \emph{edge-activation rule}, in which a uniformly random edge of the network is selected and both endpoints interact; and a \emph{node-activation rule}, in which a node is first selected uniformly at random and then one of its neighbours is drawn uniformly to interact with it. The two rules differ in how strongly an agent's degree determines its chance of being drawn into an interaction, and we use the comparison as a control in Section~\ref{speed}. The two selection rules coincide in FC and RRN because every node has the same degree (see Appendix~\ref{app:theory-rrn}). 

Each condition, defined by a model, a word pair, a topology, an activation rule and a population size, is simulated $1000$ times at $N = 1000$, $5000$ and $\num{10000}$, and $300$ times at $N = \num{100000}$. For every combination of topology and population size we generate $200$ independent network realisations ($50$ for $N = \num{100000}$). \href{https://github.com/robertopassaro/llm-naming-game-networks}{Code and data} are publicly available.

In Sections~\ref{sec:methods-degree} and~\ref{sec:mixed-state}, we use \emph{annealed networks} as a control to isolate the effects of correlations in agents' memory states caused by repeated interactions~\cite{dorogovtsev_critic_phenom_2008}. An annealed network keeps the degree of every node but rewires its edges, so that agents no longer interact persistently with the same individuals. We simulate it in the limit of infinitely fast rewiring, in which all edges are reshuffled before every interaction. In this limit the network structure, and with it every correlation of the memory states between neighbours, is lost. The only remaining property is the network degree sequence. By comparing annealed networks to the standard ``quenched'' (fixed) networks, we can determine whether an observed effect is driven by the correlations that build up between neighbours' states over time.

\subsection{Degree-resolved success on UCM}
\label{sec:methods-degree}

To resolve coordination by degree we record the number of interactions involving a degree-$k$ node and the number of them that succeeded. Their ratio is the cumulative degree-resolved success rate
\begin{equation}
    S_k = \frac{\text{successful interactions involving a degree-}k
    \text{ node}}
    {\text{total interactions involving a degree-}k \text{ node}}.
    \label{eq:Sk}
\end{equation}

$S_k$ is cumulative and increases with the population round $t$. For this analysis we considered only UCM$_{2.5}$ at $N = \num{10000}$. Since nodes with high degrees are rare, we divide the vertices into five degree bins, $k = 4$--5, 6--9, 10--17, 18--35 and 36--100, holding from $49.6\%$ down to $2.5\%$ of the population; we call the nodes in the lowest bin \emph{low-degree nodes} and the ones in the highest bin \emph{hubs}.

For each bin we then locate the crossing point $(t^{*}, n^{*})$ at which $S_k$ first reaches three target success levels: $0.6$, $0.7$ and $0.8$. Here $t^{*}$ is measured in population rounds and $n^{*}$ is the mean number of interactions a node of that bin has had by that point. The crossings of the five degree bins form an iso-success contour, which we summarise by its \emph{tilt}, $t^{*}(\text{low-degree nodes}) / t^{*}(\text{hub})$, greater than $1$ when hubs reach the level in fewer population rounds than low-degree nodes. We report the tilt on the normalised scale $\tau$ defined in Section~\ref{sec:degree}.

\subsection{Committed minorities}

To probe the robustness of an established convention against a committed minority, we seed a fixed fraction $p$ of the population with agents that always produce one target convention and never update their state, while the remaining agents begin in the steady consensus state on the competing convention, implemented by initialising them with a full memory of agreeing interactions in that convention. For each fraction $p$, we execute 200 independent runs to estimate the probability that the population flips to the committed convention within the simulation horizon, and we locate the critical fraction $p^*$, the smallest committed fraction at which the flip probability reaches $0.5$, by bisection on the committed count to a resolution of $\Delta p \leq 0.01$. Within each pair, we called \textit{strong} the convention preferred by the population in FC during convergence, and \textit{weak} the other. Both commitment directions are tested for every word pair. On the UCM networks we rank the nodes by their degree, and we place the committed set either at the top of the ranking, at the bottom, or at random.


\section{Results}

The consensus-dynamics analysis spans four population sizes, four topologies, eleven word pairs and four models under edge activation. In Section~\ref{speed} node activation is also tested, on the same population sizes and word pairs. The committed-minority analyses (Section~\ref{committed}) and the degree-resolved success analyses (Section~\ref{sec:degree}) are run for $N = \num{10000}$.

\subsection{Consensus dynamics across topologies}
\label{consensus}

Network structure can potentially change which of the two conventions the population selects. We therefore compared, across topologies, the fraction of runs in which the established convention differs from that on FC. Excluding the two mixed-state  Llama pairs \{young/old\} and \{more/less\}, $42$ model $\times$ word-pair combinations reach consensus in every run across all topologies, values of $N$, and activation rules. In $40$ of them, the same word wins most runs under all conditions. The two exceptions are GPT pairs. For \{she/he\}, \emph{she} wins $95\%$ of FC runs at $N = \num{1000}$ and all runs at larger $N$, whereas \emph{he} wins $71\%$, $95\%$, and $100\%$ of RRN runs at $N = \num{1000}$, $\num{10000}$, and $\num{100000}$, respectively (edge activation). For \{tall/short\}, \emph{tall} wins at least $99.5\%$ of FC runs, whereas \emph{short} wins $94.5$--$100\%$ on RRN and $54$--$88\%$ on UCM$_{5.0}$. Node activation yields the same pattern. In both pairs, the reversal grows with $N$, ruling out finite-size fluctuations.

\subsubsection{Speed of consensus}
\label{speed}

\begin{figure}[ht!]
\centering
\includegraphics[width=\textwidth]{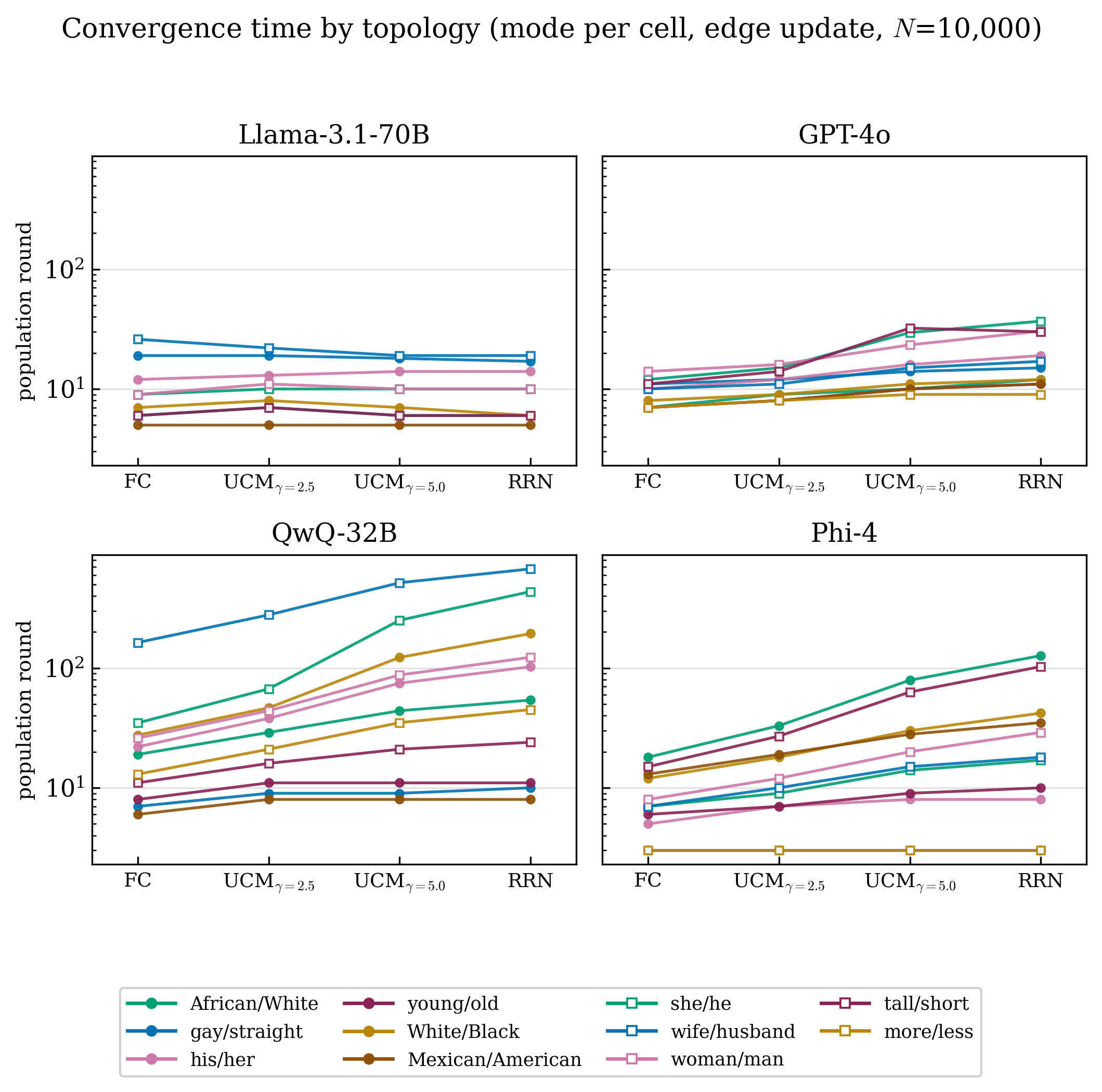}
\caption{Mode convergence time under edge activation at $N=10000$, one panel per model. Each line connects the four topology values FC, UCM$_{2.5}$, UCM$_{5.0}$, RRN; colour and marker shape jointly identify the word pair and are consistent across panels.}
\label{fig:slopegraph}
\end{figure}

To test whether topology systematically changes how fast a population reaches consensus, we compared the mode of the convergence time of every combination of word pair. We use the mode because it is less sensitive to outliers, but repeating the analysis on the mean gives the same results  (Table~\ref{tab:speed_consensus}, Figure~\ref{fig:slopegraph_mean}). 
For QwQ, GPT, and Phi, the four topologies differ significantly. Their speeds consistently rank as FC < UCM$_{2.5}$ < UCM$_{5.0}$ < RRN across all tested conditions (Figure~\ref{fig:slopegraph}; Table~\ref{tab:speed_consensus}). However, Llama does not follow this pattern for eight pairs: two have FC as the fastest topology with the rest statistically indistinguishable, three show no resolvable topology effect at all, and three entirely invert the global ordering.

Overall, denser topologies support faster consensus in three of the four models. To further investigate the role of density, we analysed the convergence time of RRN at $N = \num{10000}$ with edge activation, across increasing values of average degree (Figure~\ref{fig:density_rrn_mode}). The results, supported by statistical tests (Table~\ref{tab:density_rrn}), show that for QwQ, GPT, and Phi, RRN convergence time decreases with $K$, asymptotically approaching that of FC. For instance, at $K=4$, every one of the eleven QwQ pairs converges between $1.3$ (\{Mexican/American\}) and $12.2$ (\{she/he\}) times slower than on FC; but by $K=16$ no pair is more than $1.8$ times slower than FC, and by $K=256$ none is more than $1.05$ times slower. RRN convergence time in Llama also approaches that of FC asymptotically, but at low $K$ it can be higher or lower than FC depending on the word pair.

\subsubsection{Degree-resolved success on UCM}
\label{sec:degree}

Now we ask whether hubs on UCM$_{2.5}$ reach coordination faster than low-degree nodes, which first requires knowing how many interactions a node of degree $k$ accumulates over time.
In the symmetric edge update version of the naming game, one interaction is given by selecting an edge, and both nodes at the extremes of the edge take part in the interaction. A network of $N$ nodes with mean degree $\langle k \rangle$ has $E = N \langle k \rangle / 2$ edges. Therefore, a node of degree $k$ takes part in any single interaction with probability $k / E$. Since a population round is $N$ interactions, the expected number of interactions a node with degree $k$ has accumulated after $t$ rounds is~\cite{newman_networks_2010}
\begin{equation}
    n_k(t) \;=\; \frac{k}{E}\,N t \;=\; \frac{2 k t}{\langle k \rangle}.
    \label{eq:exposure}
\end{equation}

Because Eq.~\ref{eq:exposure} is proportional to $t$, the exposure ratio between two degree classes is $k / k'$ at every moment and never changes with time. The number of interactions of a given node is therefore a function of its degree and the population rounds.

\begin{figure}[ht!]
\centering
\includegraphics[width=\textwidth]{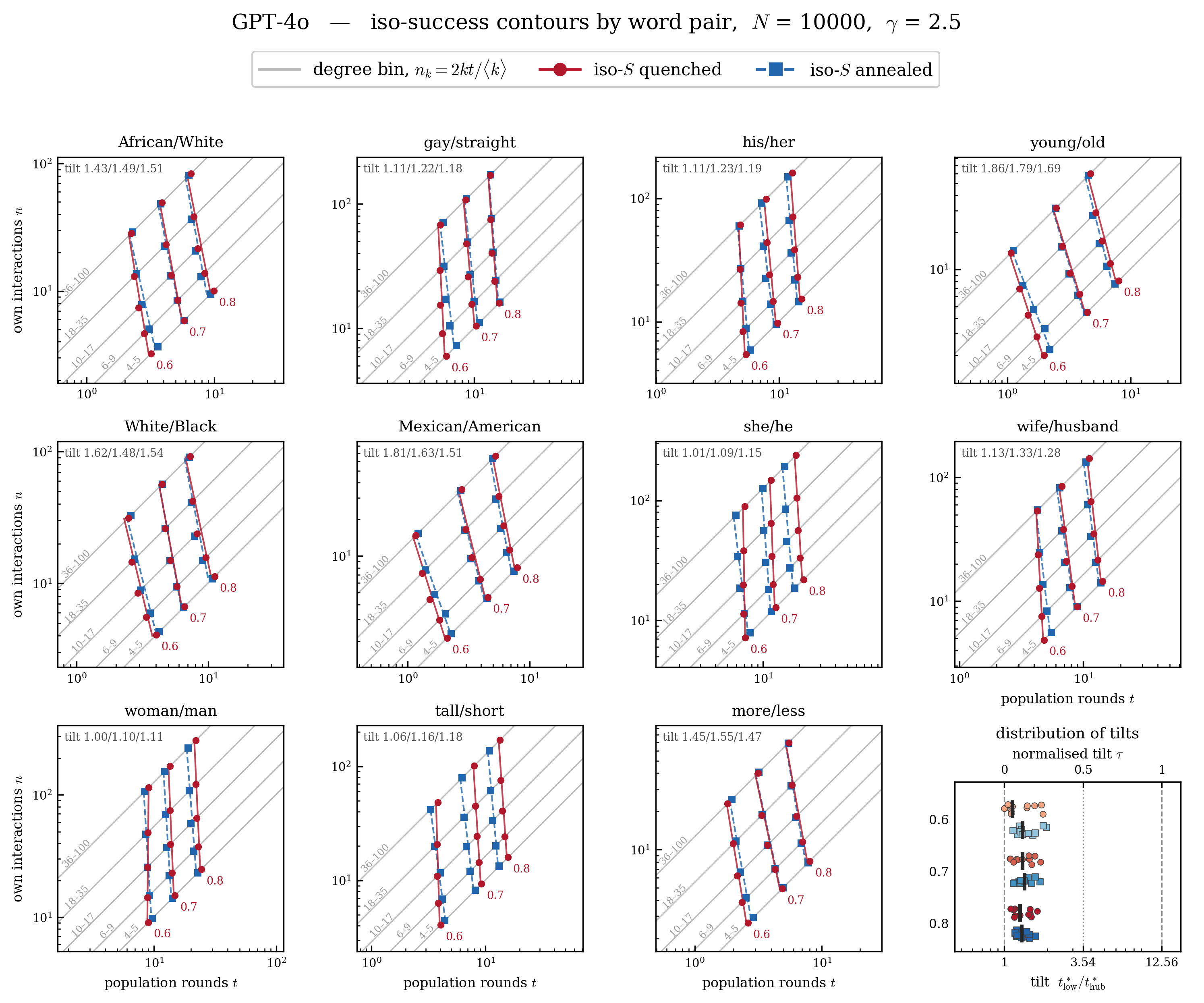}
\caption{Exposure and success by degree for GPT on UCM$_{2.5}$ at $N = 10000$, one panel per word pair. Axes are population rounds $t$ and the node's own cumulative interaction count $n$. The straight grey lines mark the range of possible values for nodes in each of the five degree bins, according to Eq.~\ref{eq:exposure}. Markers are the mean across runs at which each bin's cumulative success rate $S_k$ first reaches $0.6$, $0.7$ and $0.8$: red circles for the quenched network, blue squares for the annealed control. The line between the markers is a linear fit (solid quenched, dashed annealed). The top-left corner of each panel reports the tilt of the quenched contour at the three levels, given by the ratio $t^{*}(\text{low-degree nodes}) / t^{*}(\text{hubs})$ between the population rounds at which the lowest ($k = 4$--5) and the highest ($k = 36$--100) degree bins first reach $0.6$, $0.7$ and $0.8$. The bottom-right panel shows the distribution of the tilts across word pairs, one row per success level $S$ (labelled on the left), for the quenched (red circles) and the annealed network (blue squares), with the median as a vertical bar; the bottom axis gives the tilt, the top axis the normalised tilt $\tau$ (Eq.~\ref{eq:tau}), and the dashed lines mark the vertical ($\tau = 0$) and horizontal ($\tau = 1$) contours, the dotted line their midpoint ($\tau = 0.5$).}
\label{fig:isosuccess}
\end{figure}

To investigate the interaction between success, population rounds and a node's own interactions, we place the point at which each degree bin first reaches a given success level in the $(t, n)$ plane defined by Eq.~\ref{eq:exposure}. Figure~\ref{fig:isosuccess} shows linear regressions, each connecting the points at which the degree bins reach a given success level. Their orientation distinguishes two possible regimes. If the contours are vertical, all degree bins reach the same success level at the same population round, even though hubs have taken part in many more interactions than low-degree nodes. In this regime, success depends on the age of the population. If the contours are horizontal, each degree bin reaches the same success level after the same number of its own interactions, regardless of when this happens in population time. In this regime, success depends on individual experience. The tilt measures this orientation, and its two extreme values follow directly from Eq.~\ref{eq:exposure}. For a vertical contour $t^{*}$ (population rounds at which a given success level is reached) is the same for every bin, so the tilt is $1$. For a horizontal contour $n^{*}$ (number of interactions in which a given success level is reached) is the same for every bin, so the tilt is $k_{\text{hub}} / k_{\text{low}}$. For UCM$_{2.5}$ and $N = 10000$, this is $12.56$, the ratio between the mean degrees of the two bins ($54.81$ and $4.36$).

To make the tilt easier to read, we map it onto a normalised scale,
\begin{equation}
    \tau \;=\; \frac{\ln(\text{tilt})}{\ln(k_{\text{hub}} / k_{\text{low}})},
    \label{eq:tau}
\end{equation}
so that a tilt of $1$ (vertical contour) corresponds to $\tau = 0$ and a tilt of $12.56$ (horizontal contour) corresponds to $\tau = 1$. A negative $\tau$ means that hubs reach the level later than low-degree nodes in population time.

In Figure~\ref{fig:isosuccess} the contours are close to vertical in all eleven word pairs and at all three levels: $\tau$ of GPT goes from $-0.002$ to $0.24$. The same holds for the word pairs of the other models (Figures~\ref{fig:isosuccess_llama}, \ref{fig:isosuccess_qwq}, and~\ref{fig:isosuccess_phi}). Excluding the mixed-state conditions and those in which the initial success is above the thresholds, the normalised tilt across conditions has median $0.16$ (interquartile range $0.07$--$0.24$) and is below $0.5$ in $113$ of the $115$ conditions. The only two exceptions are Phi \{she/he\} ($0.55$) and \{woman/man\} ($0.54$), both at $S = 0.6$.

However, hubs still reach a given success level first: $\tau$ is above $0$ in $109$ of the $115$ conditions. This is consistent with the picture known for the naming game and other consensus dynamics on heterogeneous networks~\cite{dall2006nonequilibrium, castellano2009statistical}. Hubs settle on a convention early in the dynamics, and their nearest neighbours tend to share it. Moving away from the hubs this agreement is lost, and nodes remain in mixed states. The six conditions with a negative $\tau$ are all at $S = 0.6$ (Table~\ref{tab:negative_tau}). At this level the population is still disordered: a hub interacts with many neighbours holding conflicting conventions and cannot yet settle, while a low-degree node can reach an agreement with its few neighbours. All six have a positive $\tau$ at $S = 0.8$ ($0.04$--$0.15$). 

The lead of the hubs comes from the fact that they interact more: at the median $\tau = 0.16$, hubs have taken part in $\approx 8.4$ times as many interactions as low-degree nodes. This is analogous to epidemic spreading, where hubs keep the infection alive because they take part in more transmission events~\cite{pastor_review_2015}. 

To test whether the advantage of the hubs depends on correlations between the states of neighbouring agents, and not only on their higher exposure, we repeat every simulation on the annealed networks described in Section~\ref{sec:networks}. The annealed network preserves the degree of every node, but draws new edges at every interaction, so that no correlation between neighbours can build up. For each model--word pair and success level we compute the difference $\Delta\tau = \tau_{\text{annealed}} - \tau_{\text{quenched}}$, which equals $0$ when removing persistent neighbourhoods leaves the contour unchanged. At $S = 0.8$, $\Delta\tau$ has median $0.005$ (interquartile range from $-0.004$ to $0.013$). At this level, the advantage of the hubs is therefore accounted by their exposure alone. At $S = 0.6$, instead, $\Delta\tau$ is positive in $34$ of $38$ conditions (median $0.05$), because the annealed network delays $t^{*}$ of the lowest degree bin by $15\%$ and that of the highest by only $4\%$ (medians). In the six conditions in which hubs arrive last on the quenched network (Table~\ref{tab:negative_tau}), the delay is $8\%$ to $200\%$ for low-degree nodes and $-17\%$ to $+12\%$ for hubs. Therefore, in the disordered phase, a low-degree node can reach agreement with its few persistent neighbours, and this early advantage disappears once its partners are reshuffled.

Removing persistent neighbourhoods also changes the speed of consensus. For each of the 42 model $\times$ word-pair conditions in which all runs reach consensus on both networks, we compare the median convergence time on the annealed network with that on the quenched network. The annealed network is faster in 31 conditions, equally fast in 8, and slower in 3. Median convergence time on the annealed network is $14\%$ faster overall (median across conditions; range from $55\%$ faster to $27\%$ slower; Wilcoxon signed-rank test on log times, $p = 2 \times 10^{-6}$), and the reduction grows with the quenched convergence time (Spearman $\rho = -0.69$, $p < 10^{-6}$): for the ten slowest conditions (quenched median $21$-$281$ rounds), consensus time drops by $45\%$. Together with the tilt, these results suggest that persistent neighbourhoods sustain local agreements that conflict with the rest of the population. This helps low-degree nodes early on but delays global consensus, and increasingly so the slower consensus already is.

\subsubsection{Topology effect on mixed-state pairs}
\label{sec:mixed-state}

In the well-mixed population, the Llama pairs \{young, old\} and \{more, less\} reach consensus at small system size $N$ but, above a critical size, freeze in a mixed state in which neither convention ever wins~\cite{flint2026group}. Adding network structure rescues this failure, but only for the two topologies with the most homogeneous degree distributions: UCM$_{5.0}$ and RRN (Table~\ref{tab:mixed-state}). The rescue is slower on UCM$_{5.0}$, and its cost grows with $N$: the median convergence time rises from $108$ to $267$ rounds between $N = \num{1000}$ and $\num{100000}$, against $64$ to $79$ on RRN (Table~\ref{tab:mixed-state_times}). No topology rescues the pair \{more/less\} at any size tested except RRN at $N = 1000$ ($0.10$, median $508$ rounds), which fails at all larger sizes.

To investigate further the cause of the rescued consensus, we run UCM$_{5.0}$ and RRN using annealed network controls, which preserve every node's degree while removing persistent neighbourhoods. A stable pair \{african/white\} was run alongside as a negative control and reaches consensus in every condition, confirming the annealing itself is not the source of any failure to converge. Under annealing, both UCM$_{5.0}$ and RRN fail to rescue either mixed-state pair at any population size: the consensus fraction is exactly $0.00$ for \{young/old\} and \{more/less\} on both topologies at every $N \in \{1000, 5000, 10000, 100000\}$ (300 runs per condition). For RRN this is expected: in the annealed network approximation, the dynamics on an RRN of any degree $K$ reduces to FC (Appendix~\ref{app:theory}). This shows that the rescue in networks with homogeneous degree requires the dynamical correlations between the states of neighbours.

\begin{table}[ht!]
\centering
\caption{Consensus fraction for the two mixed-state Llama pairs across topologies and population sizes. 1000 rounds for condition.}
\label{tab:mixed-state}
\begin{tabular}{llcccc}
\toprule
Pair & Topology & $N=1000$ & $N=5000$ & $N=10000$ & $N=100000$ \\
\midrule
young/old & FC (well-mixed) & 0.00 & 0.00 & 0.00 & 0.00 \\
          & UCM$_{2.5}$ & 0.07 & 0.00 & 0.00 & 0.00 \\
          & UCM$_{5.0}$ & \textbf{1.00} & \textbf{1.00} & \textbf{1.00} & \textbf{1.00} \\
          & RRN & \textbf{1.00} & \textbf{1.00} & \textbf{1.00} & \textbf{1.00} \\
\midrule
more/less & FC (well-mixed) & 0.00 & 0.00 & 0.00 & 0.00 \\
          & UCM$_{2.5}$ & 0.00 & 0.00 & 0.00 & 0.00 \\
          & UCM$_{5.0}$ & 0.00 & 0.00 & 0.00 & 0.00 \\
          & RRN & 0.10 & 0.00 & 0.00 & 0.00 \\
\bottomrule
\end{tabular}
\end{table}

\begin{table}[ht!]
\centering
\caption{Mean / median critical fraction $p^*$ by placement strategy, UCM only, $N = 10000$. Both commitment directions pooled (88 values per cell).}
\label{tab:p-star-placement}
\begin{tabular}{llcc}
\toprule
Topology & Placement & Mean & Median \\
\midrule
UCM$_{2.5}$ & hub    & 0.090 & 0.026 \\
                 & random & 0.211 & 0.160 \\
                 & low-degree   & 0.366 & 0.349 \\
\midrule
UCM$_{5.0}$ & hub    & 0.145 & 0.064 \\
                 & random & 0.190 & 0.128 \\
                 & low-degree   & 0.213 & 0.154 \\
\bottomrule
\end{tabular}
\end{table}

\subsection{Committed minorities}
\label{committed}
\subsubsection{Degree-dependent placement on UCM}

\begin{figure}[p]
\centering
\begin{subfigure}{0.85\textwidth}
  \centering
  \includegraphics[width=\textwidth]{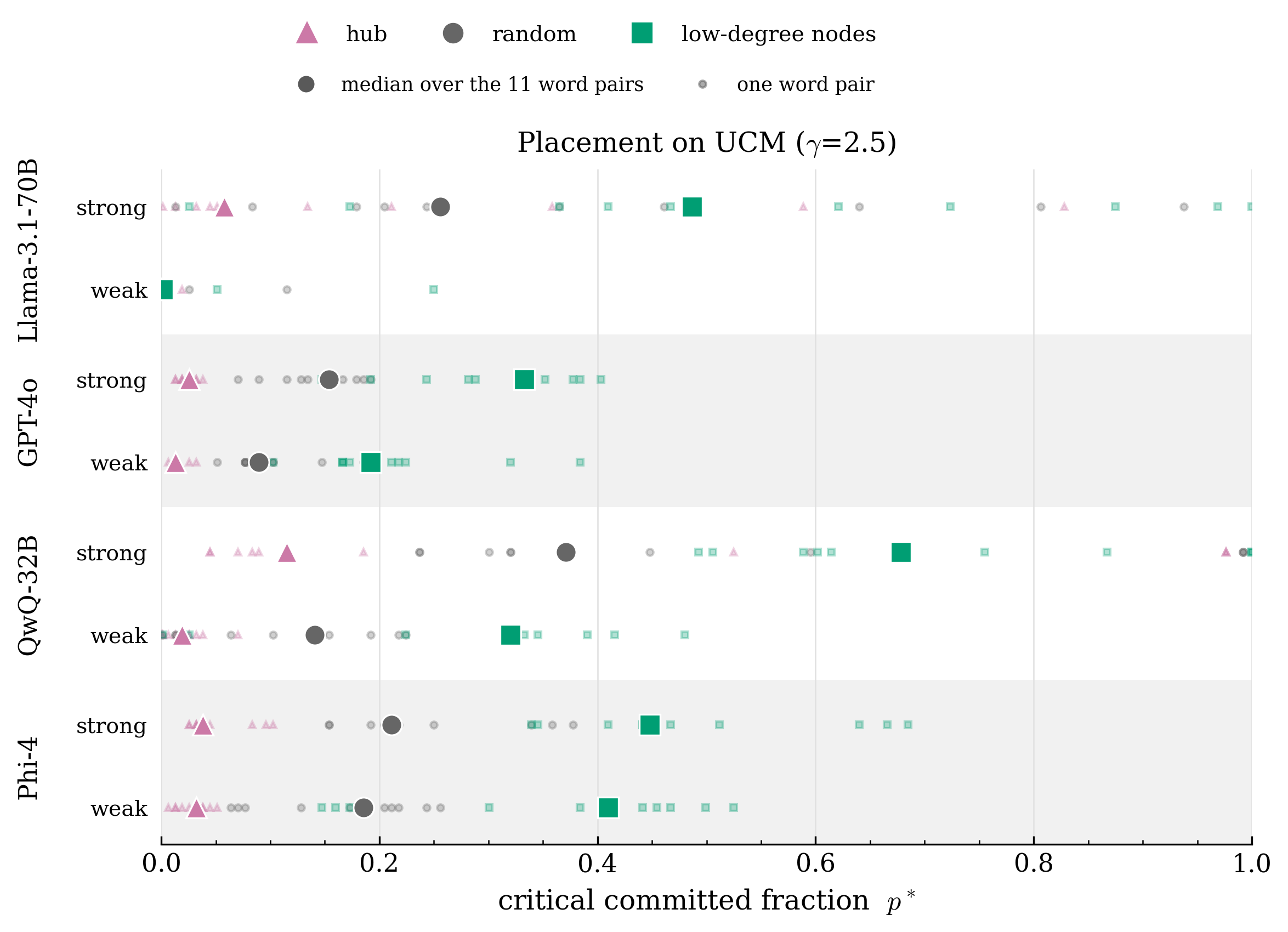}
  \caption{Placement (hub / random / low-degree nodes) on UCM$_{2.5}$).}
  \label{fig:rq2b}
\end{subfigure}

\vspace{0.8\baselineskip}

\begin{subfigure}{0.85\textwidth}
  \centering
  \includegraphics[width=\textwidth]{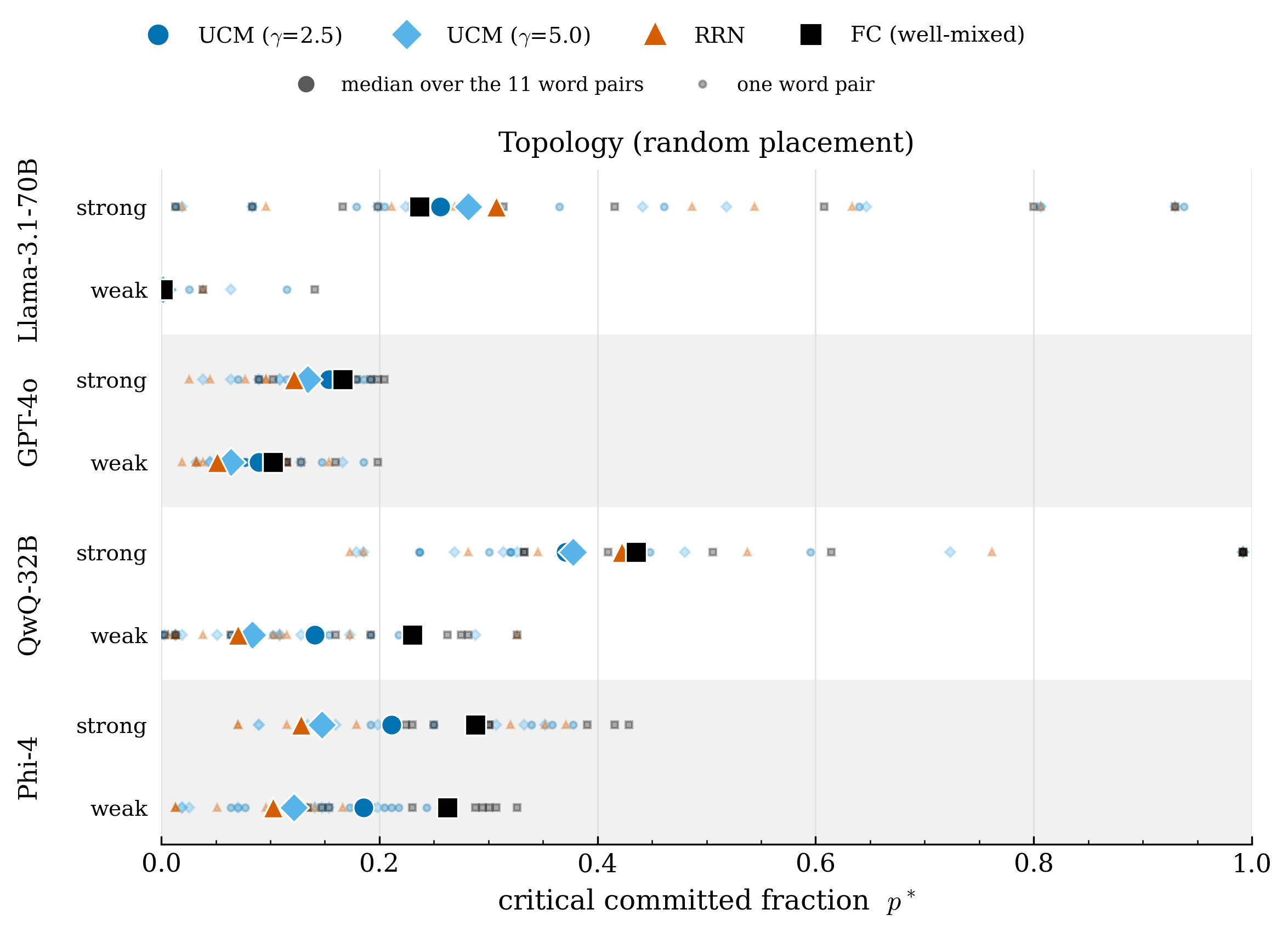}
  \caption{Topology (FC, UCM$_{2.5}$, UCM$_{5.0}$, RRN) under random placement.}
  \label{fig:rq2a}
\end{subfigure}

\caption{Critical committed fraction $p^*$ at $N = 10000$, by model and commitment direction. \textbf{(a)} $p^*$ by placement of the committed minority on UCM$_{2.5}$. \textbf{(b)} $p^*$ by topology under random placement. In both panels, large symbols are the median over the 11 word pairs and small symbols the individual pairs; the $p^* = 0$ conditions are excluded from the plot.}
\label{fig:rq2}
\end{figure}

On a heterogeneous network, the ability of a committed minority to overturn an established convention may depend not only on its size but also on the position of its members. We therefore compare the critical fraction $p^*$ on UCM when the committed agents occupy the hubs, randomly chosen nodes, or the low-degree nodes. The ordering $p^*(\text{hub}) \leq p^*(\text{random}) \leq p^*(\text{low-degree nodes})$ holds in all conditions on both UCM exponents (Figure~\ref{fig:rq2b}; Table~\ref{tab:p-star-placement}). To test whether this ordering is systematic, we compared $p^*$ of every word pair across the three placements, separately for each model, direction and UCM exponent (Table~\ref{tab:rq2_placement}).
The three placements differ significantly for every model and direction on both exponents ($W \geq 0.85$), with the ranks ordered as hub $<$ random $<$ low-degree nodes. The only exception is Llama in the weak direction (the one not preferred when reaching consensus on FC). There, in six of the nine pairs the weak convention is overturned by $1$ to $32$ committed agents out of $\num{10000}$, below the resolution of the bisection: the resulting ties make the tests non-significant, although the ordering holds in all nine pairs. The degree exponent changes the size of the placement effect: the gap $p^*(\text{low-degree nodes}) - p^*(\text{hub})$ is larger on $\gamma = 2.5$ than on the near-homogeneous $\gamma = 5.0$ in 83 of the 86 pairs, as expected since $83.6\%$ of nodes lie at $k \leq 5$ on $\gamma = 5.0$, against $49.5\%$ on $\gamma = 2.5$.

In 15 conditions $p^* = 0$: the established consensus does not hold even with zero committed agents. These are systematic: all Llama, on the pairs \{her, his\} and \{short, tall\}, always in the direction starting from consensus on \emph{her} and \emph{short}, and present on every topology and placement (both UCMs at all three placements, RRN, and FC for \{short, tall\}). Because $p^* = 0$ denotes the absence of a convention to overturn rather than a small critical mass, these conditions are omitted from Figures~\ref{fig:rq2b} and~\ref{fig:rq2a}. At the opposite extreme, 27 conditions require $p^* \geq 0.95$, and in 7 of them it is exactly $1.00$, meaning the convention flips only when the entire population is committed. All seven occur under low-degree nodes placement.

\subsubsection{Topology effects on the critical mass}

The weak and strong conventions of a given pair exhibit unequal robustness: displacing the strong convention requires a larger $p^*$ than displacing the weak one in 34 of the 44 pairs on FC (4 ties), and under random placement in 37 on UCM$_{2.5}$ and 38 on both UCM$_{5.0}$ and RRN. This reproduces the directional asymmetry reported by \cite{flint2026group}.

To test whether topology systematically changes the critical mass, we compared $p^*$ of every word pair under random placement across FC, UCM$_{2.5}$, UCM$_{5.0}$ and RRN, separately for each model and direction (Figure~\ref{fig:rq2a}; Table~\ref{tab:pstar_topology} reports the median $p^*$ of each combination). 
For GPT and Phi the four topologies differ significantly in both directions, with the ranks in the predicted order; the same trend holds for QwQ in the weak direction (Table~\ref{tab:rq2_topology}).
There are three exceptions. For Llama in the strong direction the four topologies differ significantly ($W = 0.64$), but the ordering is \emph{reversed}: RRN requires the largest committed minority and FC the smallest (rank sums $38.5$, $32.5$, $25.0$, $14.0$ for RRN, UCM$_{5.0}$, UCM$_{2.5}$ and FC). For QwQ in the strong direction the four topologies differ ($W = 0.32$, $p < .05$) but without a trend (Page's $L$ not significant). Finally, for Llama in the weak direction topology has no significant effect ($p = .11$). Therefore, the predicted ordering is supported for three models in the weak direction and for two in the strong direction.

\begin{table}[ht!]
\centering
\caption{Median critical fraction $p^*$ over the word pairs, by model, commitment direction and topology under random placement, $N = 10000$. Strong and weak as in Figure~\ref{fig:rq2}; the $p^* = 0$ conditions are excluded. Statistical tests are in Table~\ref{tab:rq2_topology}.}
\label{tab:pstar_topology}
\begin{tabular}{llcccc}
\toprule
Model & Direction & RRN & UCM$_{5.0}$ & UCM$_{2.5}$ & FC \\
\midrule
Llama-3.1-70B & strong & 0.307 & 0.282 & 0.256 & 0.237 \\
              & weak   & 0.003 & 0.002 & 0.001 & 0.002 \\
\midrule
GPT-4o        & strong & 0.122 & 0.134 & 0.154 & 0.166 \\
              & weak   & 0.051 & 0.064 & 0.090 & 0.102 \\
\midrule
QwQ-32B       & strong & 0.422 & 0.378 & 0.371 & 0.435 \\
              & weak   & 0.070 & 0.083 & 0.141 & 0.230 \\
\midrule
Phi-4         & strong & 0.128 & 0.147 & 0.211 & 0.288 \\
              & weak   & 0.102 & 0.122 & 0.186 & 0.262 \\
\bottomrule
\end{tabular}
\end{table}

\section{Discussion}

We embedded populations of LLM agents playing the minimal naming game in networks of controlled degree structure and measured how they converge on a convention and how easily a committed minority can overturn it. Three main results emerge.

First, denser topologies support faster consensus: for most word pairs, convergence time is ordered FC $<$ UCM$_{2.5}$ $<$ UCM$_{5.0}$ $<$ RRN. Because they interact more often, hubs reach a given success level slightly earlier than low-degree nodes, consistent with classical naming game experiments in which agreement forms around hubs and spreads outward~\cite{dall2006nonequilibrium}. Persistent neighbourhoods matter mainly in the disordered phase, where low-degree nodes can settle local agreements with their few fixed partners, echoing the known tension between locally stabilising structure and global convergence~\cite{baronchelli2006topology, lu2009naming, mistry2015committed}. These results parallel human systems, where topology is known to affect convergence~\cite{baronchelli2006topology, dall2006nonequilibrium, castellano2009statistical}.

Second, topology occasionally changes which convention wins (Section~\ref{consensus}) and whether consensus emerges at all, rescuing a pair that never converges in a well-mixed population. The rescue vanishes when neighbours are reshuffled, showing that dynamical correlations between persistent neighbours are necessary for it. Unlike density leading to faster consensus, these effects have not been observed in human populations.

Third, the critical fraction $p^*$ needed to overturn an established convention follows $p^*(\text{hub}) \leq p^*(\text{random}) \leq p^*(\text{low-degree nodes})$ in all conditions. The most cost-effective attack on a multi-agent LLM system is thus to target its hubs~\cite{chan2024visibility, chan2025infrastructure}, mirroring degree-targeted attacks~\cite{albert2000error}. Unlike prompt injection, jailbreaks, or manipulated knowledge propagating between agents~\cite{gu2024agent, lee2025prompt, ju2026flooding}, this attack compromises neither the agents nor the content they exchange: it merely places ordinary, persistently opinionated agents at well-connected positions. For randomly placed committed minorities, the ordering RRN $<$ UCM$_{5.0}$ $<$ UCM$_{2.5}$ $<$ FC holds in only five of the eight model-direction combinations tested, highlighting variability across models. In the minimal naming game, the tipping point is known to shift with mean degree and connectivity~\cite{zhang2012analytic}, and committed agents at central or highly active positions are far more effective than randomly placed ones~\cite{lu2009naming, mistry2015committed}. We show that these effects persist, and sharpen, in LLMs.

Our work has limitations. First, results are model dependent: Llama departs from the majority pattern in speed ordering, in the topology dependence of $p^*$, and in having mixed-state pairs, so more models are needed before generalising. Second, the minimal naming game makes collective properties testable at population sizes beyond the reach of direct LLM querying, but it is a simplified task; future work should test more ecological scenarios with richer interactions~\cite{park2023generative, perez2024cultural, nohr2026peer}. Third, we fixed memory at $H = 5$ and the number of competing conventions at $W = 2$. Consensus in LLM populations depends on memory~\cite{mehdizadeh2026exploring}, and larger convention sets open indirect routes between equilibria that change the cost of an attack~\cite{flint2026indirect}.

Despite these limitations, our results bear directly on the safety of multi-agent systems, where heterogeneous topologies are increasingly common~\cite{ehtesham2025survey, chan2025infrastructure, price2026let}. In most of our settings, well-mixed populations reach consensus faster and require larger committed minorities to be overturned than networked ones, so well-mixed estimates can overstate robustness. In a few cases, topology even changes which convention is established and whether consensus is reached. Moreover, because all agents in a well-mixed population are equivalent, the effect of committed-agent placement cannot be observed there. This matters because the conventions agent populations adopt can have tangible consequences, from prices set by autonomous pricing agents~\cite{calvano2020artificial, fish2024algorithmic} to narratives amplified by agent swarms~\cite{schroeder2026malicious}. The safety of multi-agent systems depends on these collective outcomes as much as on individual model behaviour~\cite{hammond2025multi, brinkmann2023machine}. As agents are increasingly deployed in networks, assessing their safety will require mapping who interacts with whom, alongside testing the agents themselves.

\paragraph{Funding:} L.M.A. acknowledges the support from the Horizon-CL4 project MINGLE (EU grant: 101341932), and from the Carlsberg Foundation through the COCOONS project (CF21-0432). R.P.-S. acknowledges financial support from project PID2022-137505NB-C21/AEI/10.13039/501100011033/FEDER UE and the Acadèmia d’Excel·lència Award, funded by the Generalitat de Catalunya.

\bibliography{references}

\clearpage

\appendix
\setcounter{section}{0}
\setcounter{figure}{0}
\setcounter{table}{0}
\setcounter{equation}{0}
\renewcommand{\thesection}{A\arabic{section}}
\renewcommand{\thetable}{A\arabic{table}}
\renewcommand{\thefigure}{A\arabic{figure}}
\renewcommand{\theequation}{A\arabic{equation}}
\renewcommand{\theHsection}{A\arabic{section}}
\renewcommand{\theHtable}{A\arabic{table}}
\renewcommand{\theHfigure}{A\arabic{figure}}
\renewcommand{\theHequation}{A\arabic{equation}}


\section*{APPENDIX}

\section{Mean-field theory of the naming game on networks}
\label{app:theory}

\subsection{Mean-field naming game for LLMs}

We consider a population composed by agents (LLMs) playing the naming game with
the same vocabulary (same words $0$ and $1$).  The state of the agents is given
by the set of words proposed and received in the previous interactions. Thus, a
state can be represented as a string of bits of the form
\begin{equation}
  \label{eq:1}
  S = \{a_1, b_1, a_2, b_2,  \ldots, a_h, b_h\},
\end{equation}
where $h \leq H$ is the length of the state, $a_i$ are the last words proposed by the agent and $b_i$ are the last words received by interacting partners. $H$ is the maximum length of the state, corresponding to the maximum memory of the agents. Agents start from the empty state $S_\emptyset = \emptyset$ (empty memory stack before any interaction). The total number of states is 
\begin{equation}
  M = \frac{1}{3} (4^{H + 1} - 1)
\end{equation}

Agents are characterised by a probability policy $q(S)$ defined as the
probability that an agent in state $S$ will select the word $0$ as output of an
interaction with another agent (irrespective of the state of the second agent),
selecting word $1$ with probability $1 - q(S))$. At the mean field level (everybody interacting with everybody else), the dynamics of the system is given by the following rules:
\begin{enumerate}
\item Two agents are randomly selected to interact.
\item The first agent, in state $S$, proposes a word $a$ according to its probability policy $q(S)$.
\item The second agent, in state $S'$, receives the word $a$ and responds with a word $b$ according to its probability policy $q(S')$.
\item The states of both agents are updated by adding the pair words uttered and received to their respective histories.
\end{enumerate}

If an agent in state $S$ proposes the word $a$ and receives the word $b$, its state is updated  according to the following rule:
\begin{equation}
  \label{eq:2}
  S  \to \Gamma_{(a, b)}(S),
\end{equation}
where the shift function $\Gamma_{(a,b)}(S)$ is defined as
\begin{equation}
  \label{eq:3}
  \Gamma_{(a,b)}( \{a_1, b_1, a_2, b_2,  \ldots, a_h, b_h\}) = \left \{
    \begin{array}{ll}
      \{a_1, b_1, a_2, b_2,  \ldots, a_h, b_h, a, b\}  &
                                                              \;\mathrm{if} \; h
                                                              < H \\
                                                            & \\
      \{a_2, b_2, a_3, b_3,  \ldots, a_H, b_H, a, b\}  &
                                                              \;\mathrm{if} \; h
                                                              = H \\
    \end{array}
  \right. .
\end{equation}

At the mean field level, the state of the system is given by the state vector $\vec{x}$, where $x_k$ is the fraction of agents in state $k$. The state vector is normalised as  $\sum_k  x_k = 1$, such that the state of the system is represented by a point in a $M$-dimensional simplex. The state vectors fulfill the following rate equations:
\begin{equation}
  \frac{d x_k}{dt} =  -2x_k +  2\sum_i \sum_j x_i x_j P(k | i, j),
\end{equation}
where the kernel $P(k | i, j)$, the probability that an agent in state $i$ will transition to state $k$ when interacting with an agent in state $j$, is computed in terms of the possible transitions of $i$ and the probability policy $q(i)$ as:
\begin{equation}
    P(k | i, j) = \left \{
  \begin{array}{ll}
    q(i) q(j) & \mathrm{if} \; k =
                                          \Gamma_{(0,0)}(i),
    \\
    q(i) (1-q(j)) & \mathrm{if} \; k =
                                              \Gamma_{(0,1)}(i),
    \\
    (1 - q(i)) q(j) & \mathrm{if} \; k =
                                              \Gamma_{(1,0)}(i),
    \\
    (1 - q(i)) (1 - q(j)) & \mathrm{if} \; k =
                                                \Gamma_{(1,1)}(i),
    \\
    0 & \mathrm{otherwise}
    \end{array}
  \right. ,
  \label{eq:kernel}
\end{equation}
Notice that we have kept a factor of $2$ in the rate equation, usually absorbed in a time rescaling, in order to keep the same time scale as in the networked version of the model.

\subsection{The naming game in complex networks}

We consider now a population of $N$ agents placed on the nodes of a complex network. The network is characterised by its adjacency matrix $A_{nm}$~\cite{newman_networks_2010}, where $A_{nm} = 1$ if nodes $n$ and $m$ are connected by means of an edge and $A_{nm} = 0$ otherwise. The degree of node $n$ is given by $k_n = \sum_m A_{nm}$, representing the number of nearest neighbours directly connected to node $n$.

We consider two different update dynamics:
\begin{enumerate}
  \item \textbf{Symmetric node update:} We choose at random a node $n$ in the network, and then we choose at random one of its nearest neighbours $m$. Both nodes propose a word according to their respective state. We update the state of both nodes $n$ and $m$.
  \item \textbf{Edge update:} We choose at random an edge $(n, m)$ in the network. Both nodes propose a word according to their respective state. We update the state of both nodes $n$ and $m$.
\end{enumerate}

We assume that the probability policies are not affected by the network structure, in such a way that the update of any node depends only on its own state and the state of node with which it is interacting. 

\subsection{Quenched mean-field theory}

In order to proceed to a mathematical formulation, we consider the so-called quenched mean-field theory (QMF)~\cite{chakrabarti_qmf_2008,pastor_review_2015}, in which we represent the state of each node $n$ in the network with a  variable $x_i^n$, representing the probability that the agent placed in node $n$ is in state $i$. Since each agent is in one of the $M$ possible states, we have the normalisation condition 
\begin{equation}
  \sum_i x_i^n = 1, \quad \forall n = 1, 2, \ldots, N.
\end{equation}
That is, the state of each node is represented by a point in the $M$-dimensional simplex. The mean-field approximation consists in assuming that the state of each node is independent of the state of its neighbours, in such a way that the probability that a node $n$ in state $i$ and another node $m$ in state $j$ are joined by and edge, $P(n,i ; m,j)$  is given by 
\begin{equation}
  P(n,i ; m,j) = P(n,i) P(m,j) = x_i^n x_j^m.
\end{equation}
Notice that this lack of dynamical correlations is an approximation, since the state of a node is correlated with the state of its neighbours, due to the interactions between them. It can be improved by considering higher-order dynamical correlations in a pair approximation~\cite{mata2013pair}.

Since the state-transition kernel is independent of the network structure, the gain-loss contribution associated with an interaction between nodes $n$ and $m$ has the structure~\cite{flint2026group}:
\begin{equation}
  - x_\ell^n + \sum_{i,j} x_i^n x_j^m P(\ell | i, j),
  \label{eq:structure}
\end{equation}
The effects of topology are incorporated in the mean-field equations by considering the different probabilities to choose nodes $n$ and $m$ in the network. Let us consider the different update rules in turn.

\subsubsection{Symmetric node update}

At each elementary time step, we choose a node $n$, with probability $1/N$, that will be in state $i$ with probability $x_i^n$. We then select a nearest neighbour $m$ of node $n$ with probability $A_{nm}/k_n$, where $k_n$ is the degree of node $n$. Combining these probabilities with the gain-loss structure in Eq.~\eqref{eq:structure}, we obtain the following mean-field equations for symmetric node update:
\begin{eqnarray}
  \frac{d x_\ell^n}{dt} &=&  \frac{1}{N} \sum_m \frac{A_{nm}}{k_n} \left[ 
      - x_\ell^n + \sum_{i,j} x_i^n x_j^m P(\ell | i, j) \right]  + \frac{1}{N} \sum_m \frac{A_{nm}}{k_m} \left[ 
      - x_\ell^n + \sum_{i,j} x_i^n x_j^m P(\ell | i, j) \right] \nonumber \\
      & = & \frac{1}{N} \left[
        - x_\ell^n \left( 1 + \sum_m \frac{A_{nm}}{k_m} \right) + \sum_m A_{nm} \left( \frac{1}{k_n} + \frac{1}{k_m} \right) \sum_{i,j} x_i^n x_j^m P(\ell | i, j)
      \right].
\end{eqnarray}
The two terms on the rhs of the first equation represent the  contributions from the node $n$ and its neighbours, respectively. Absorbing a factor of $1/N$ in a time rescaling, we can write the mean-field equations for the symmetric node update as
\begin{equation} 
  \frac{d x_\ell^n}{dt} =  - x_\ell^n \left( 1 + \sum_m \frac{A_{nm}}{k_m} \right) + \sum_m A_{nm} \left( \frac{1}{k_n} + \frac{1}{k_m} \right) \sum_{i,j} x_i^n x_j^m P(\ell | i, j).
\end{equation}

\subsubsection{Symmetric edge update}

At each elementary time step, we choose an edge $(n, m)$ in the network with probability $A_{nm}/E$, where $E = \sum_{n,m} A_{nm}/2$ is the total number of edges in the network. The gain-loss structure in Eq.~\eqref{eq:structure} leads to the following mean-field equations for the symmetric edge update:
\begin{equation}
  \frac{d x_\ell^n}{dt} =  \sum_m \frac{A_{nm}}{E} \left[ 
      - x_\ell^n + \sum_{i,j} x_i^n x_j^m P(\ell | i, j) \right].  
\end{equation}
Absorbing a factor of $1/N$ in a time rescaling, and considering that $E = N \langle{k}\rangle/2$, where $\langle{k}\rangle$ is the average degree of the network, we can write the mean-field equations for the symmetric edge update as
\begin{equation}
  \frac{d x_\ell^n}{dt} = \frac{2}{\langle{k}\rangle} \left[ -k_n  x_\ell^n +  \sum_m A_{nm} \sum_{i,j} x_i^n x_j^m P(\ell | i, j) \right]
\end{equation}  

\subsection{Annealed network approximation}

The annealed network approximation~\cite{dorogovtsev_critic_phenom_2008,boguna_langevin_approach_2009} consists in replacing the adjacency matrix $A_{nm}$ by its expected value, given by the connection probability $p_{nm}$ between nodes $n$ and $m$. For an uncorrelated configuration-model network, in the sparse-network limit, we replace the adjacency matrix by~\cite{boguna_langevin_approach_2009}
\begin{equation}
  A_{nm} \to p_{nm} = \frac{k_n k_m}{2 E} = \frac{k_n k_m}{\langle{k}\rangle N},
\end{equation}
where $\langle{k}\rangle$ is the average degree of the network. An annealed network is a network in which the edges are randomly rewired, at a rate much faster than the dynamics of the nodes, while keeping the degree of each node fixed. The annealed network approximation is interesting because, due to the edge rewiring, it renders exact the hypothesis of lack of dynamical correlations between the states of neighbouring nodes, which is an approximation in the quenched mean field theory. Thus, we expect that numerical simulations in annealed networks will be exactly described by the annealed network approximation, while in quenched networks we can have deviations from the theory, either QMF or the annealed network approximation, due to the presence of dynamical correlations. In any case, the annealed network approximation  renders simplex equations that sometimes capture the scaling behaviour of the dynamics.

Within the annealed network approximation, the mean-field equations for the symmetric node update reduce to
\begin{equation}
  \frac{d x_\ell^n}{dt} =  -x_\ell^n \left( 1 + \frac{k_n}{\langle{k}\rangle}  \right) + \sum_m \frac{k_n + k_m}{\langle{k}\rangle N}  \sum_{i,j} x_i^n x_j^m P(\ell | i, j),
\end{equation}
while the equations for the symmetric edge update reduce to
\begin{equation}
  \frac{d x_\ell^n}{dt} = \frac{2}{\langle{k}\rangle} \left[ -k_n  x_\ell^n +  \sum_m \frac{k_n k_m}{\langle{k}\rangle N} \sum_{i,j} x_i^n x_j^m P(\ell | i, j) \right]
\end{equation} 

\subsection{Random regular networks}
\label{app:theory-rrn}

Consider a random regular network (RRN), in which all nodes have the same degree $k_n = K$, that is $\sum_m A_{nm} = K$ for all nodes $n$. In this case, the QMF equations for the symmetric node update reduce to
\begin{equation}
  \frac{d x_\ell^n}{dt} =  - 2 x_\ell^n  + \frac{2}{K}  \sum_m A_{nm} \sum_{i,j} x_i^n x_j^m P(\ell | i, j),
\end{equation}
while the equations for the symmetric edge update reduce to
\begin{equation}
  \frac{d x_\ell^n}{dt} =  - 2 x_\ell^n  + \frac{2}{K}  \sum_m A_{nm} \sum_{i,j} x_i^n x_j^m P(\ell | i, j).
\end{equation}
Thus, for random regular networks, the QMF equations for both update rules are identical.

Within the annealed network approximation, the equations for both symmetric node update and symmetric edge update reduce to
\begin{equation}
  \frac{d x_\ell^n}{dt} =  - 2 x_\ell^n  + \frac{2}{N}  \sum_m  \sum_{i,j} x_i^n x_j^m P(\ell | i, j),
\end{equation}
which is independent of the degree $K$ of the network. Thus, for RRN, the annealed network approximation predicts that the dynamics is independent of the degree of the network.

Moreover, if we define the average state vector $\rho_i = \sum_n x_i^n/N$, we can write the mean-field equations for the average state vector as
\begin{equation}
  \frac{d \rho_\ell}{dt} =  - 2 \rho_\ell  + 2 \sum_{i,j} \rho_i \rho_j P(\ell | i, j),
\end{equation}
which is identical to the mean-field equations for the fully connected network. Thus, the annealed network approximation predicts that the dynamics of the naming game in RRN is identical to the dynamics in the fully connected network for any value of $K$.

\clearpage

\section{Additional figures and tables}

In Tables \ref{tab:speed_consensus}, \ref{tab:density_rrn}, 
\ref{tab:rq2_placement}, and \ref{tab:rq2_topology}, we present the results of Friedman and Page tests. The Friedman test asks whether the four topologies differ in convergence time, comparing them within each word pair; Page's $L$ asks whether they follow the specific ordering FC $<$ UCM$_{2.5}$ $<$ UCM$_{5.0}$ $<$ RRN.

\subsection{Statistical tests for speed of consensus across topologies}

\begin{table}[ht!]
\centering
\caption{Friedman and Page's $L$ tests on the mode of reached consensus of the topology ordering
FC~$<$~UCM$_{2.5}$~$<$~UCM$_{5.0}$~$<$~RRN. Performed by model, activation rule and population size.}
\label{tab:speed_consensus}
\begin{tabular}{lllccccc}
\toprule
Model & $N$ & Rule & $Q$ & $p$ & $W$ & $L$ & $p(L)$ \\
\midrule
QwQ-32B & 1000    & Edge & 31.4 & $<.001$ & 0.953 & 327 & $<.001$ \\
QwQ-32B & 1000    & Node & 31.4 & $<.001$ & 0.953 & 327 & $<.001$ \\
QwQ-32B & 5000    & Edge & 30.7 & $<.001$ & 0.930 & 326 & $<.001$ \\
QwQ-32B & 5000    & Node & 31.4 & $<.001$ & 0.950 & 327 & $<.001$ \\
QwQ-32B & 10000   & Edge & 30.7 & $<.001$ & 0.930 & 326 & $<.001$ \\
QwQ-32B & 10000   & Node & 31.4 & $<.001$ & 0.950 & 327 & $<.001$ \\
QwQ-32B & 100000  & Edge & 30.7 & $<.001$ & 0.930 & 326 & $<.001$ \\
QwQ-32B & 100000  & Node & 32.2 & $<.001$ & 0.975 & 328 & $<.001$ \\
\midrule
GPT-4o  & 1000    & Edge & 32.7 & $<.001$ & 0.992 & 330 & $<.001$ \\
GPT-4o  & 1000    & Node & 32.7 & $<.001$ & 0.992 & 330 & $<.001$ \\
GPT-4o  & 5000    & Edge & 31.9 & $<.001$ & 0.967 & 329 & $<.001$ \\
GPT-4o  & 5000    & Node & 31.7 & $<.001$ & 0.962 & 328 & $<.001$ \\
GPT-4o  & 10000   & Edge & 31.7 & $<.001$ & 0.962 & 328 & $<.001$ \\
GPT-4o  & 10000   & Node & 31.9 & $<.001$ & 0.967 & 329 & $<.001$ \\
GPT-4o  & 100000  & Edge & 31.4 & $<.001$ & 0.953 & 328 & $<.001$ \\
GPT-4o  & 100000  & Node & 31.0 & $<.001$ & 0.938 & 328 & $<.001$ \\
\midrule
Phi-4   & 1000    & Edge & 26.7 & $<.001$ & 0.810 & 320 & $<.001$ \\
Phi-4   & 1000    & Node & 26.7 & $<.001$ & 0.810 & 320 & $<.001$ \\
Phi-4   & 5000    & Edge & 26.7 & $<.001$ & 0.810 & 320 & $<.001$ \\
Phi-4   & 5000    & Node & 26.7 & $<.001$ & 0.810 & 320 & $<.001$ \\
Phi-4   & 10000   & Edge & 26.7 & $<.001$ & 0.810 & 320 & $<.001$ \\
Phi-4   & 10000   & Node & 26.7 & $<.001$ & 0.810 & 320 & $<.001$ \\
Phi-4   & 100000  & Edge & 26.7 & $<.001$ & 0.810 & 320 & $<.001$ \\
Phi-4   & 100000  & Node & 26.7 & $<.001$ & 0.810 & 320 & $<.001$ \\
\midrule
Llama-3.1-70B & 1000    & Edge & 1.7 & 0.644 & 0.062 & 224 & 0.568 \\
Llama-3.1-70B & 1000    & Node & 0.8 & 0.857 & 0.028 & 224 & 0.568 \\
Llama-3.1-70B & 5000    & Edge & 5.9 & 0.114 & 0.220 & 222 & 0.655 \\
Llama-3.1-70B & 5000    & Node & 1.0 & 0.801 & 0.037 & 224 & 0.612 \\
Llama-3.1-70B & 10000   & Edge & 7.5 & 0.058 & 0.277 & 220 & 0.734 \\
Llama-3.1-70B & 10000   & Node & 1.0 & 0.801 & 0.037 & 224 & 0.612 \\
Llama-3.1-70B & 100000  & Edge & 9.0 & 0.029 & 0.335 & 218 & 0.804 \\
Llama-3.1-70B & 100000  & Node & 1.3 & 0.737 & 0.047 & 224 & 0.612 \\
\bottomrule
\end{tabular}
\end{table}

\begin{figure}[ht!]
\centering
\includegraphics[width=\textwidth]{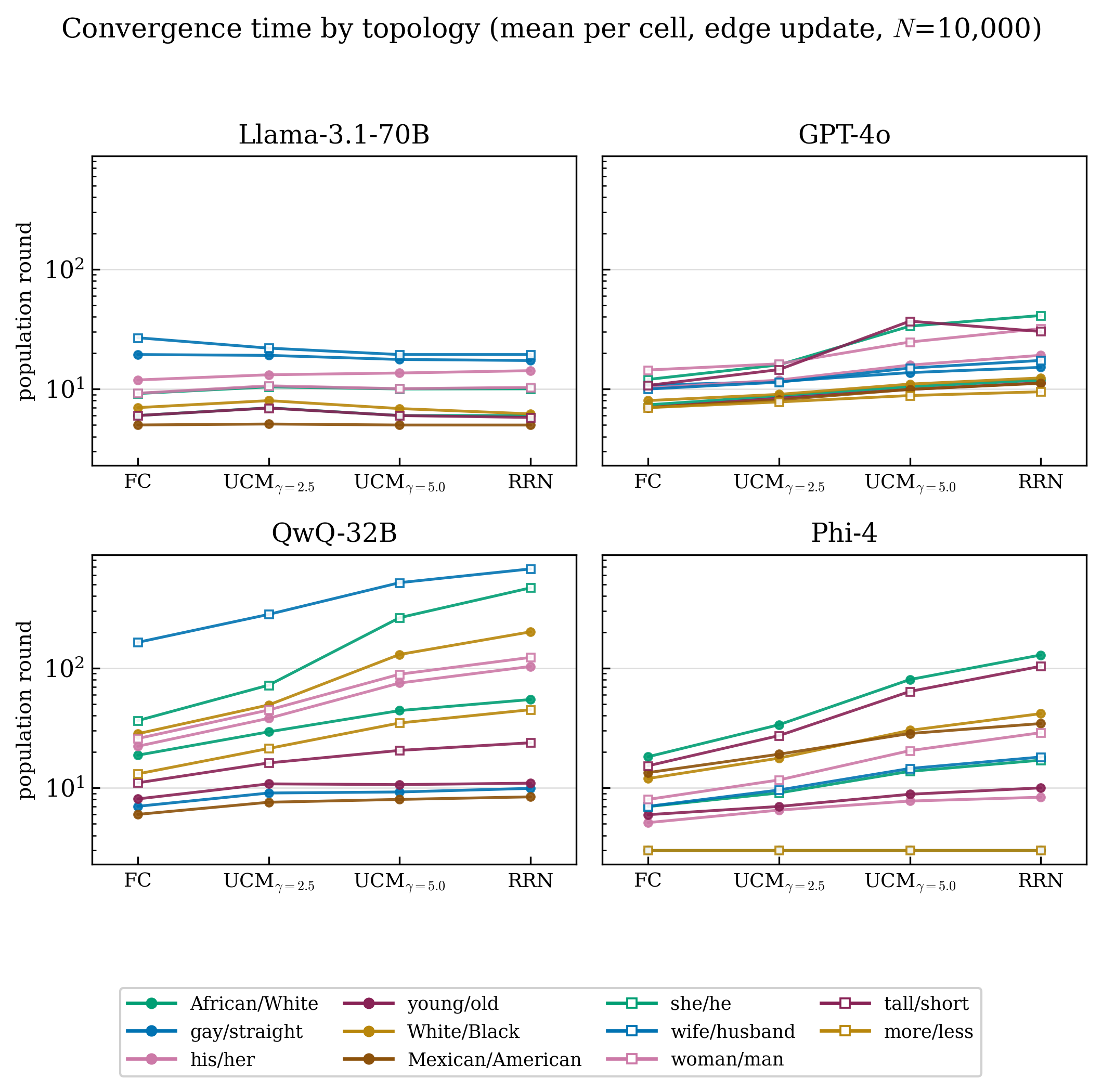}
\caption{Same as Figure~\ref{fig:slopegraph}, but for the mean convergence time under edge activation at $N=10000$.}
\label{fig:slopegraph_mean}
\end{figure}

\clearpage

\subsection{Statistical tests for speed of consensus on increasing values of K for RRN}

\begin{table}[ht!]
\centering
\caption{Friedman and Page's $L$ tests of convergence time on RRN as a function of mode degree $K \in \{4,6,8,10,12,16,20,24,32,48,64,96,128,192,256\}$ against the FC baseline, $N=10000$, edge activation. $n=11$ word pairs for QwQ-32B, GPT-4o and Phi-4; $n=9$ for Llama-3.1-70B, whose two mixed-state pairs (\S\ref{sec:mixed-state}) are excluded.}
\label{tab:density_rrn}
\begin{tabular}{lcccccccc}
\toprule
Model & $n$ & $Q$ & $p$ & $W$ & $L$ & $p(L)$ \\
\midrule
QwQ-32B       & 11 & 154.8 & $<.001$   & 0.938 & 16170 & $<.001$   \\
GPT-4o        & 11 & 138.0 & $<.001$   & 0.837 & 15456 & $<.001$   \\
Phi-4         & 11 & 123.4 & $<.001$   & 0.748 & 15427 & $<.001$   \\
Llama-3.1-70B & 9  & 2.0   & 1.000 & 0.015 & 10420 & 0.477 \\
\bottomrule
\end{tabular}
\end{table}

\clearpage

\subsection{Speed of consensus on increasing values of K for RRN}

\begin{figure}[ht!]
\centering
\includegraphics[width=\textwidth]{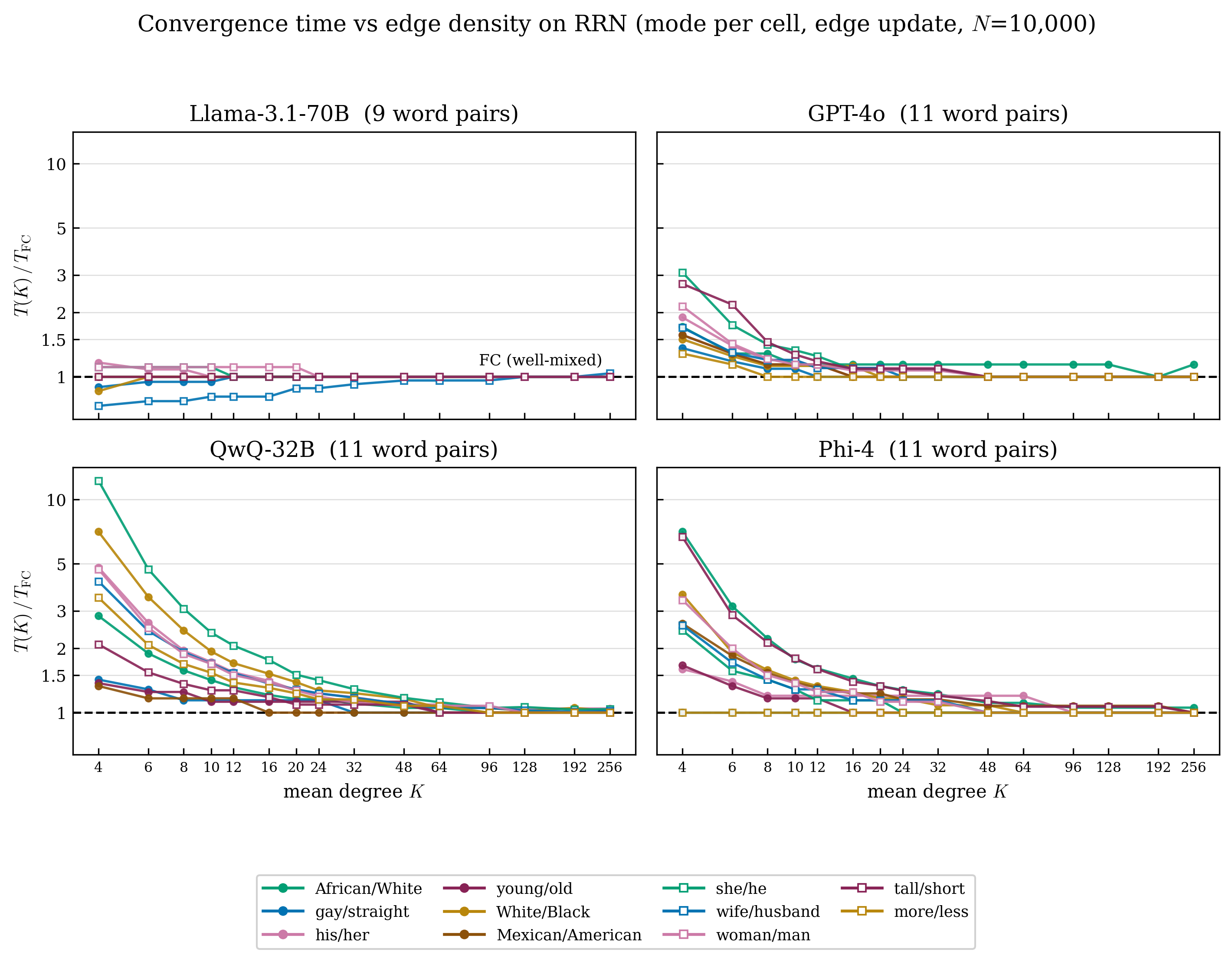}
\caption{Convergence time on random regular networks (RRN) as a function of the degree $K$, relative to the fully connected (FC) population. For every model--word-pair combination, we take the mode of the convergence time of RRN at degree $K$, $T(K)$, and divide it by $T_\mathrm{FC}$. A ratio above $1$ means the network is slower than FC. The dashed line at $1$ marks the FC baseline. $N=10000$, edge activation, $K \in \{4,6,8,10,12,16,20,24,32,48,64,96,128,192,256\}$.}
\label{fig:density_rrn_mode}
\end{figure}

\clearpage

\subsection{Convergence time at increasing N for the mixed-state Llama pairs}

\begin{figure}[ht!]
\centering
\includegraphics[width=\textwidth]{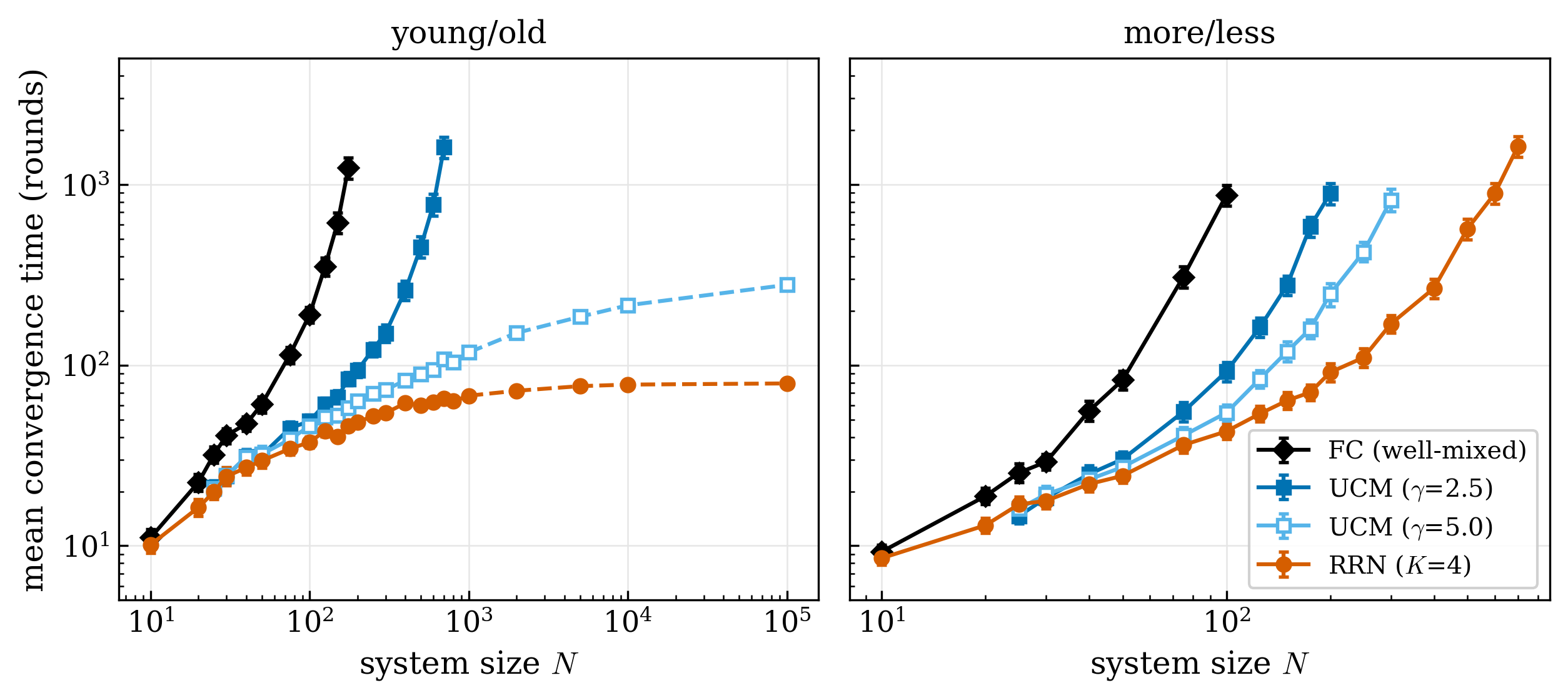}
\caption{Mean convergence time as a function of $N$ for the two mixed-state Llama-3.1-70B pairs \{young/old\} (left) and \{more/less\} (right), edge activation, on FC (black diamonds), UCM$_{2.5}$, UCM$_{5.0}$ (blue open squares) and RRN with $K = 4$ (orange circles). Error bars are $95\%$ bootstrap confidence intervals over $200$ runs per point, with a fresh network per run. Runs last at most $t_\mathrm{max} = 10000$ rounds. Each curve starts at $N = 10$ ($N = 25$ for UCM$_{2.5}$ and UCM$_{5.0}$), and stops at the last $N$ at which all runs reach consensus within $t_\mathrm{max}$.}
\label{fig:mixed-state_N}
\end{figure}

\clearpage

\subsection{Convergence time on RRN at increasing K for the mixed-state Llama pairs}

\begin{figure}[ht!]
\centering
\includegraphics[width=\textwidth]{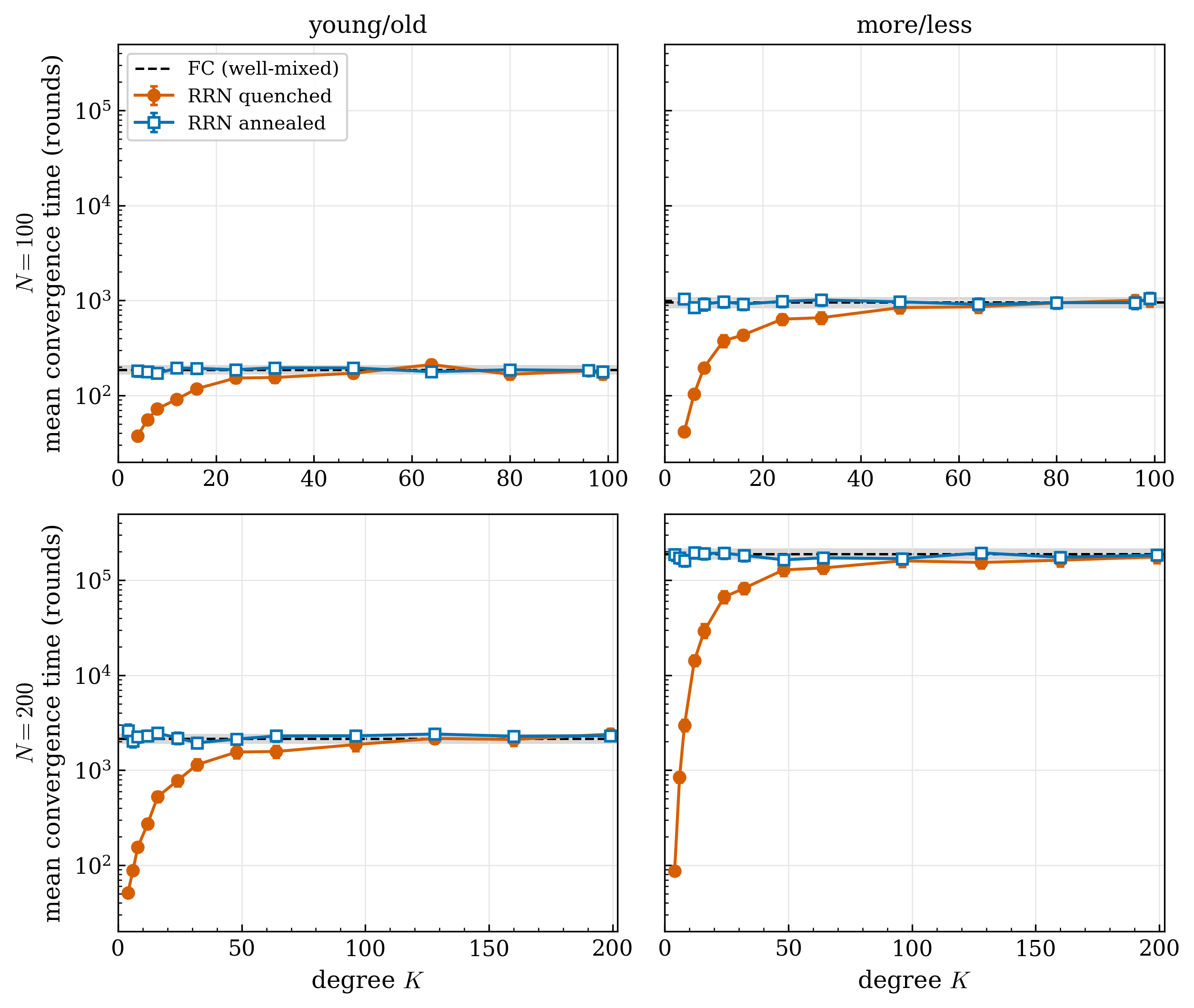}
\caption{Mean convergence time on RRN as a function of the degree $K$, for the two mixed-state Llama-3.1-70B pairs \{young/old\} (left) and \{more/less\} (right), at $N = 100$ (top) and $N = 200$ (bottom), edge activation. Orange circles are quenched networks, blue open squares the annealed control; error bars are $95\%$ bootstrap confidence intervals over $200$ runs per condition. The dashed line and the grey band give the FC mean and its confidence interval at the same $N$. Note the logarithmic vertical axis, and that the rightmost point of each panel, $K = N-1$, is the fully connected graph.}
\label{fig:mixed-state_K}
\end{figure}

\clearpage

\begin{table}[ht!]
\centering
\caption{Median convergence time (population rounds) of the two mixed-state Llama-3.1-70B pairs, edge activation; consensus fraction in parentheses when below $1$. A dash marks conditions in which no run reaches consensus. Consensus fractions for every topology are in Table~\ref{tab:mixed-state}.}
\label{tab:mixed-state_times}
\begin{tabular}{llcccc}
\toprule
Pair & Topology & $N=\num{1000}$ & $N=\num{5000}$ & $N=\num{10000}$ & $N=\num{100000}$ \\
\midrule
young/old & UCM$_{2.5}$ & 593.5 (0.07) & -- & -- & -- \\
          & UCM$_{5.0}$ & 108 & 167 & 193.5 & 267 \\
          & RRN         & 64  & 75  & 77    & 79 \\
\midrule
more/less & RRN         & 508 (0.10) & -- & -- & -- \\
\bottomrule
\end{tabular}
\end{table}

\begin{table}[ht!]
\centering
\caption{RRN degree sweep for the two mixed-state Llama-3.1-70B pairs, edge activation, $200$ runs per condition. Tested degrees: $K \in \{4, 6, 8, 12, 16, 24, 32, 48, 64, 80, 96, 99\}$ at $N = 100$ and $K \in \{4, 6, 8, 12, 16, 24, 32, 48, 64, 96, 128, 160, 199\}$ at $N = 200$. $T(4)$ and $T_\mathrm{FC}$ are mean convergence times (population rounds) at $K = 4$ and on FC. $K_{50}$ and $K_{90}$ are the degrees at which $50\%$ and $90\%$ of the increase from $\log T(4)$ to $\log T_\mathrm{FC}$ is covered. $K_\mathrm{FC}$ is the smallest $K$ from which the RRN mean stays inside the $95\%$ confidence interval of FC.}
\label{tab:mixed-state_K}
\begin{tabular}{llrrrrrr}
\toprule
Pair & $N$ & $T(4)$ & $T_\mathrm{FC}$ & $T_\mathrm{FC}/T(4)$ & $K_{50}$ & $K_{90}$ & $K_\mathrm{FC}$ \\
\midrule
young/old & 100 & 37.5 & 186.8   & 5.0  & 10.5 & 36.3 & 80 \\
more/less & 100 & 41.7 & 954.5   & 22.9 & 8.1  & 35.7 & 48 \\
young/old & 200 & 51.1 & 2160    & 42.3 & 13.2 & 45.6 & 128 \\
more/less & 200 & 87.0 & 188700  & 2170 & 8.8  & 34.3 & 199 \\
\bottomrule
\end{tabular}
\end{table}

\clearpage

\subsection{Iso-success contours for Llama, QwQ and Phi}
\label{app:isosuccess}

\begin{table}[ht!]
\centering
\caption{The six conditions in which hubs reach the success level later than low-degree nodes on the quenched UCM$_{2.5}$ ($\tau < 0$ at $S = 0.6$), $N = \num{10000}$. $\Delta t^{*}$ is the relative delay of $t^{*}$ on the annealed network with respect to the quenched one, at $S = 0.6$, for the lowest ($k = 4$--5) and highest ($k = 36$--100) degree bins.}
\label{tab:negative_tau}
\begin{tabular}{llrrrr}
\toprule
Model & Pair & $\tau$ ($S = 0.6$) & $\tau$ ($S = 0.8$) & $\Delta t^{*}_{\text{low}}$ & $\Delta t^{*}_{\text{hub}}$ \\
\midrule
QwQ-32B & White/Black  & $-0.18$  & $0.08$ & $+108\%$ & $-17\%$ \\
QwQ-32B & she/he       & $-0.27$  & $0.08$ & $+160\%$ & $-6\%$ \\
QwQ-32B & wife/husband & $-0.18$  & $0.14$ & $+200\%$ & $+12\%$ \\
QwQ-32B & woman/man    & $-0.03$  & $0.15$ & $+111\%$ & $+5\%$ \\
Phi-4   & African/White& $-0.03$  & $0.15$ & $+98\%$  & $+4\%$ \\
GPT-4o  & woman/man    & $-0.002$ & $0.04$ & $+8\%$   & $-7\%$ \\
\bottomrule
\end{tabular}
\end{table}

\begin{figure}[ht!]
\centering
\includegraphics[width=\textwidth]{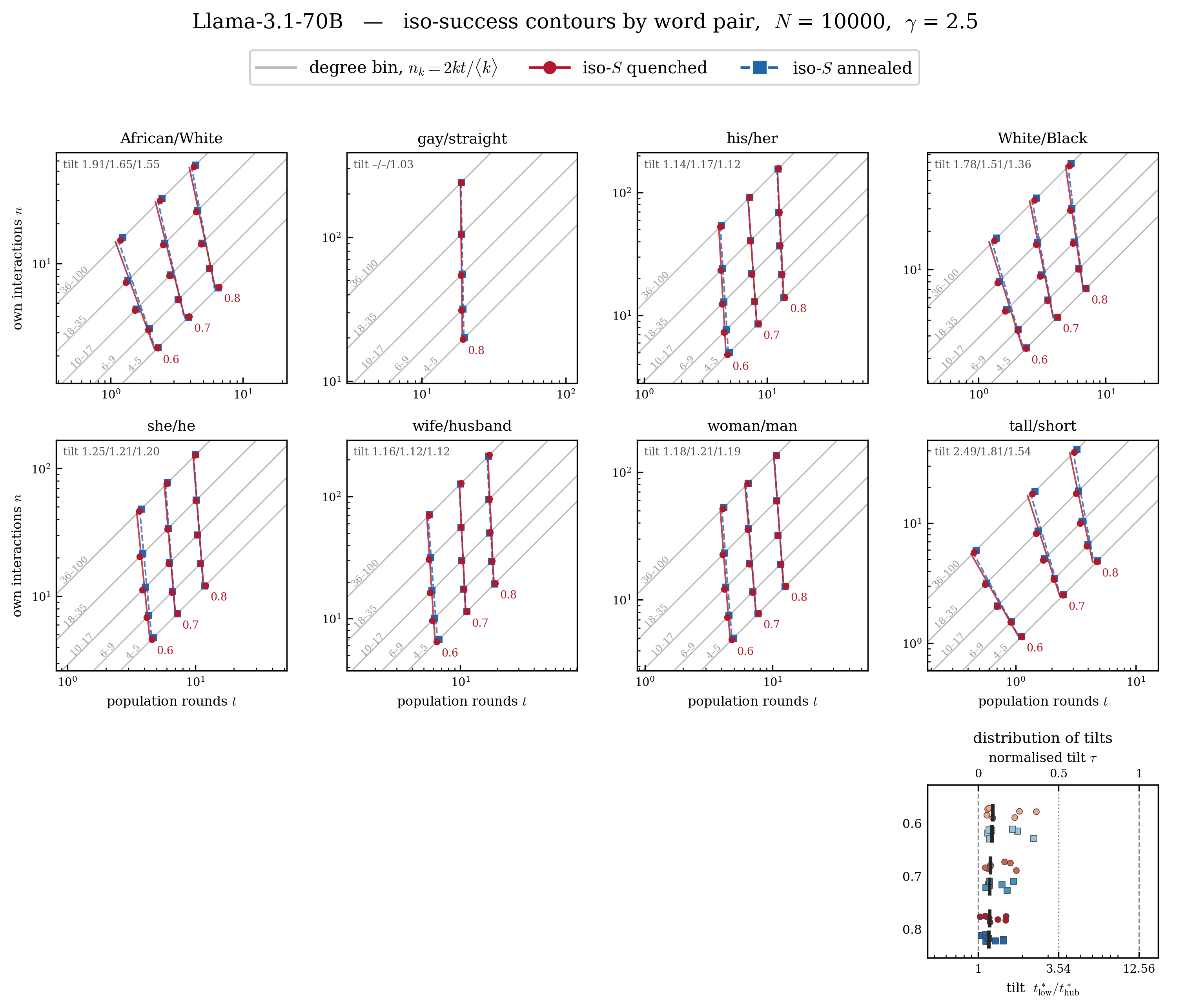}
\caption{Exposure and success by degree for Llama-3.1-70B on UCM$_{2.5}$ at $N = 10000$, one panel per word pair. Axes, markers and the bottom-right panel as in Figure~\ref{fig:isosuccess}; the top-left corner of each panel reports the quenched tilt at $S = 0.6$, $0.7$ and $0.8$, with a dash for a level that the word pair does not cross. The mixed-state pairs \{young/old\} and \{more/less\} are omitted, as is \{Mexican/American\}, whose initial success ($0.82$) lies above all three levels; for \{gay/straight\} (initial success $0.76$) only $S = 0.8$ is shown.}
\label{fig:isosuccess_llama}
\end{figure}

\clearpage

\begin{figure}[ht!]
\centering
\includegraphics[width=\textwidth]{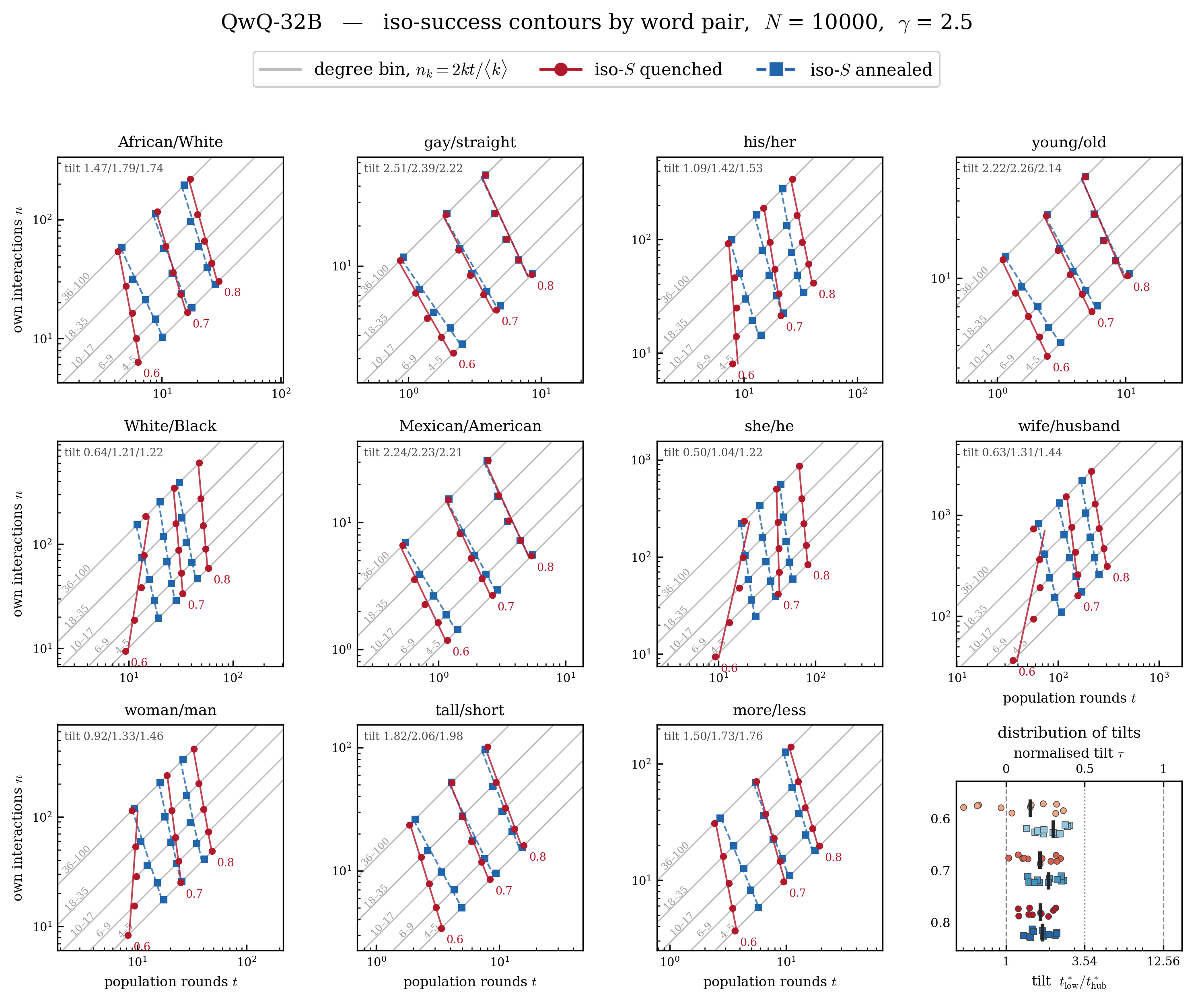}
\caption{Exposure and success by degree for QwQ-32B on UCM$_{2.5}$ at $N = 10000$, one panel per word pair. Axes, markers and the bottom-right panel as in Figure~\ref{fig:isosuccess}; the top-left corner of each panel reports the quenched tilt at $S = 0.6$, $0.7$ and $0.8$, with a dash for a level that the word pair does not cross.}
\label{fig:isosuccess_qwq}
\end{figure}

\clearpage

\begin{figure}[ht!]
\centering
\includegraphics[width=\textwidth]{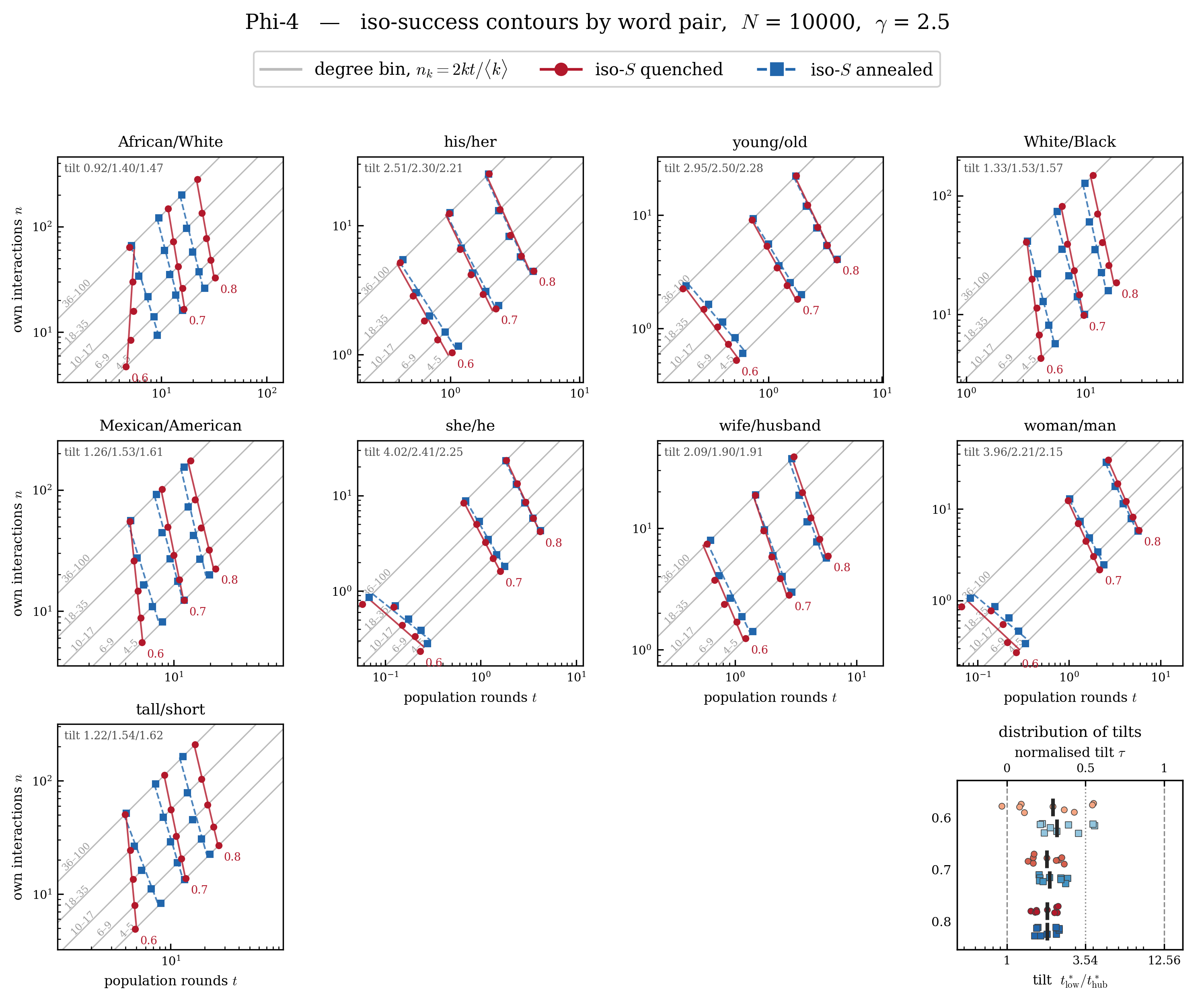}
\caption{Exposure and success by degree for Phi-4 on UCM$_{2.5}$ at $N = 10000$, one panel per word pair. Axes, markers and the bottom-right panel as in Figure~\ref{fig:isosuccess}; the top-left corner of each panel reports the quenched tilt at $S = 0.6$, $0.7$ and $0.8$, with a dash for a level that the word pair does not cross. The pairs \{gay/straight\} and \{more/less\} are omitted, since their initial success ($1.00$ and $0.94$) lies above all three levels.}
\label{fig:isosuccess_phi}
\end{figure}

\clearpage

\subsection{Statistical tests for committed minorities}

\begin{table}[ht!]
\centering
\caption{Friedman and Page's $L$ tests on the critical fraction $p^*$ of the placement ordering hub~$<$~random~$<$~low-degree nodes, by UCM exponent, model and commitment direction, $N = 10000$. The block is the word pair; $p^*$ is rounded to the bisection resolution ($0.01$) before ranking. Pairs with $p^* = 0$ in any placement are excluded ($n = 9$ for Llama-3.1-70B in the weak direction).}
\label{tab:rq2_placement}
\begin{tabular}{llcccccc}
\toprule
Model & Direction & $n$ & $Q$ & $p$ & $W$ & $L$ & $p(L)$ \\
\midrule
\multicolumn{8}{l}{\emph{UCM$_{2.5}$}} \\
\midrule
QwQ-32B       & strong & 11 & 22.0 & $<.001$ & 1.000 & 154 & $<.001$ \\
              & weak   & 11 & 20.0 & $<.001$ & 0.909 & 152 & $<.001$ \\
\midrule
GPT-4o        & strong & 11 & 22.0 & $<.001$ & 1.000 & 154 & $<.001$ \\
              & weak   & 11 & 22.0 & $<.001$ & 1.000 & 154 & $<.001$ \\
\midrule
Phi-4         & strong & 11 & 22.0 & $<.001$ & 1.000 & 154 & $<.001$ \\
              & weak   & 11 & 22.0 & $<.001$ & 1.000 & 154 & $<.001$ \\
\midrule
Llama-3.1-70B & strong & 11 & 22.0 & $<.001$ & 1.000 & 154 & $<.001$ \\
              & weak   &  9 & 5.6 & 0.060 & 0.313 & 114 & 0.147 \\
\midrule
\multicolumn{8}{l}{\emph{UCM$_{5.0}$}} \\
\midrule
QwQ-32B       & strong & 11 & 18.9 & $<.001$ & 0.858 & 150.5 & $<.001$ \\
              & weak   & 11 & 20.1 & $<.001$ & 0.916 & 152 & $<.001$ \\
\midrule
GPT-4o        & strong & 11 & 22.0 & $<.001$ & 1.000 & 154 & $<.001$ \\
              & weak   & 11 & 21.5 & $<.001$ & 0.979 & 154 & $<.001$ \\
\midrule
Phi-4         & strong & 11 & 22.0 & $<.001$ & 1.000 & 154 & $<.001$ \\
              & weak   & 11 & 21.1 & $<.001$ & 0.961 & 153 & $<.001$ \\
\midrule
Llama-3.1-70B & strong & 11 & 21.5 & $<.001$ & 0.979 & 154 & $<.001$ \\
              & weak   &  9 & 3.7 & 0.156 & 0.206 & 112 & 0.281 \\
\bottomrule
\end{tabular}
\end{table}

\begin{table}[ht!]
\centering
\caption{Friedman and Page's $L$ tests on the critical fraction $p^*$ of the topology ordering RRN~$<$~UCM$_{5.0}$~$<$~UCM$_{2.5}$~$<$~FC under random placement, by model and commitment direction, $N = 10000$. The block is the word pair; $p^*$ is rounded to the bisection resolution ($0.01$) before ranking. Pairs with $p^* = 0$ on any topology are excluded ($n = 9$ for Llama-3.1-70B in the weak direction). Page's $L$ is one-sided in the predicted direction, so $p(L) = 1.000$ for Llama-3.1-70B in the strong direction reflects the reversed ordering.}
\label{tab:rq2_topology}
\begin{tabular}{llcccccc}
\toprule
Model & Direction & $n$ & $Q$ & $p$ & $W$ & $L$ & $p(L)$ \\
\midrule
QwQ-32B       & strong & 11 & 10.5 & 0.015 & 0.319 & 286 & 0.138 \\
              & weak   & 11 & 13.1 & 0.004 & 0.396 & 305 & $<.001$ \\
\midrule
GPT-4o        & strong & 11 & 32.4 & $<.001$ & 0.982 & 329 & $<.001$ \\
              & weak   & 11 & 33.0 & $<.001$ & 1.000 & 330 & $<.001$ \\
\midrule
Phi-4         & strong & 11 & 30.4 & $<.001$ & 0.921 & 327 & $<.001$ \\
              & weak   & 11 & 33.0 & $<.001$ & 1.000 & 330 & $<.001$ \\
\midrule
Llama-3.1-70B & strong & 11 & 21.1 & $<.001$ & 0.640 & 234 & 1.000 \\
              & weak   &  9 & 6.0 & 0.112 & 0.222 & 235 & 0.139 \\
\bottomrule
\end{tabular}
\end{table}

\begin{table}[ht!]
\centering
\caption{Critical fraction $p^*$ for the three QwQ-32B pairs whose convention is overturned only by a near-complete committed minority, by topology and placement, $N = \num{10000}$. For each pair, the first row is the direction requiring $p^* \geq 0.95$, the second the opposite direction.}
\label{tab:extreme_pstar}
\begin{tabular}{lcccccccc}
\toprule
 & FC & RRN & \multicolumn{3}{c}{UCM$_{2.5}$} & \multicolumn{3}{c}{UCM$_{5.0}$} \\
\cmidrule(lr){4-6}\cmidrule(lr){7-9}
Pair & random & random & hub & random & low & hub & random & low \\
\midrule
gay/straight & 0.992 & 0.992 & 0.977 & 0.992 & 1.000 & 0.992 & 0.992 & 1.000 \\
             & 0.013 & 0.013 & 0.001 & 0.013 & 0.019 & 0.003 & 0.013 & 0.013 \\
\midrule
young/old    & 0.992 & 0.992 & 0.977 & 0.992 & 1.000 & 0.992 & 0.992 & 1.000 \\
             & 0.003 & 0.006 & $<0.001$ & 0.001 & 0.002 & 0.001 & 0.003 & 0.006 \\
\midrule
wife/husband & 0.992 & 0.992 & 0.977 & 0.992 & 1.000 & 0.992 & 0.992 & 1.000 \\
             & 0.275 & 0.326 & 0.070 & 0.218 & 0.390 & 0.250 & 0.288 & 0.314 \\
\bottomrule
\end{tabular}
\end{table}

\clearpage

\end{document}